\documentclass[times]{elsarticle}
\usepackage[utf8]{inputenc}
\usepackage[T1]{fontenc}
\usepackage{lmodern}
\usepackage{microtype}
\usepackage{hyperref}
\usepackage{url}
\usepackage{doi}
\usepackage[x11names,table]{xcolor}
\usepackage{textcomp}
\usepackage{manyfoot}
\usepackage{listings}
\usepackage{setspace}
\usepackage{lineno}
\usepackage{ragged2e}

\usepackage{graphicx}
\usepackage{booktabs}
\usepackage{longtable}
\usepackage{array}
\usepackage{multirow}
\usepackage{caption}
\usepackage{subcaption}

\usepackage{mathtools}
\usepackage{amssymb}
\usepackage{amsmath}
\usepackage{amsfonts}
\usepackage{amsthm}
\usepackage{centernot}
\usepackage{nicefrac}

\usepackage[margin=1in]{geometry}

\usepackage[capitalize]{cleveref}
\crefname{appendix}{}{}

\newcommand{\T}{\mathcal{T}}
\newcommand{\C}{\mathcal{C}}
\renewcommand{\H}{\mathcal{H}}
\newcommand{\set}[1]{\{#1\}}

\definecolor{mygreen}{rgb}{0.0,0.7,0.3}

\journal{Decision Support Systems}

\begin{document}
\begin{frontmatter}

\title{An Interactive Decision-Support Dashboard for Optimal Hospital Capacity Management}

\author[case,malone]{Felix Parker} % fparker9@jhu.edu
\author[pucv,malone,jhmi]{Diego A. Martínez} % dmart101@jh.edu
\author[jhmi]{James J. Scheulen} % jscheule@jhmi.edu
\author[case,malone]{Kimia Ghobadi\ead{kimia@jhu.edu}}

\affiliation[case]{organization={Department of Civil and Systems Engineering, The Center for Systems Science and Engineering},
            addressline={3400 N. Charles Street}, 
            city={Baltimore},
            postcode={21218}, 
            state={Maryland},
            country={USA}}
\affiliation[malone]{organization={The Malone Center for Engineering in Healthcare},
            addressline={3400 N. Charles Street}, 
            city={Baltimore},
            postcode={21218},
            state={Maryland},
            country={USA}}
\affiliation[jhmi]{organization={Department of Emergency Medicine, Johns Hopkins University School of Medicine},
            addressline={1800 Orleans St}, 
            city={Baltimore},
            postcode={21287}, 
            state={Maryland},
            country={USA}}
\affiliation[pucv]{organization={School of Industrial Engineering, Pontificia Universidad Católica de Valparaíso},
            addressline={2147 Brasil Avenue}, 
            city={Valparaíso},
            country={Chile}}

%\cortext[cor]{Corresponding author.}
%% Abstract
\begin{abstract}
Data-driven optimization models have the potential to significantly improve hospital capacity management, particularly during demand surges, when effective allocation of capacity is most critical and challenging. However, models alone cannot improve capacity management, they must be tightly integrated into existing human-led decision-making processes to realize their potential. As hospital administrators are ultimately responsible for capacity management decisions, building trustworthy and accessible tools for them is essential to effectively leverage data-driven optimization in practice. In this study, we develop an interactive, user-friendly, electronic dashboard for informing hospital capacity management decisions during surge periods. The dashboard integrates real-time hospital data, predictive analytics, and modular optimization models. It allows hospital administrators to interactively customize parameters and model components, enabling them to explore a range of scenarios, and provide real-time updates on recommended optimal decisions. The dashboard was created through a participatory design process, involving hospital administrators in the development team to ensure practical utility, trustworthiness, transparency, explainability, and usability. We successfully deployed our dashboard within the Johns Hopkins Health System during the height of the COVID-19 pandemic, addressing the increased need for tools to inform hospital capacity management. It was used on a daily basis, with results regularly communicated to hospital leadership. This study demonstrates the practical application of a prospective, data-driven, interactive decision-support tool for hospital system capacity management.
\end{abstract}

%Research highlights
% \begin{highlights}
%     \item Built interactive dashboard to support capacity management in hospitals demand surges
%     \item Dashboard uses prediction and optimization models to recommend optimal decisions
%     \item Dashboard implemented and used within a hospital system during the COVID-19 pandemic
%     \item A participatory design process was used to ensure trustworthiness and usability
% \end{highlights}

%% Keywords
% \begin{keyword}
% Dashboard Systems \sep Decision Support Systems, Management \sep Surge Capacity \sep Hospital Bed Capacity \sep COVID-19
% \end{keyword}

\begin{keyword}
    Hospital capacity management \sep Hospital surge capacity \sep Online dashboard \sep Interactive dashboard \sep Operations research \sep COVID-19
\end{keyword}

\end{frontmatter}
% \linenumbers

\section{Introduction}
\label{sec:intro}

Managing hospital capacity and utilization when demand is uncertain, resources are limited, and costs need to be controlled is a complex, challenging problem \citep{green2005}. Hospitals must balance financial incentives to maximally use available capacity with the need to preserve some capacity for potential demand surges. Managing this balance becomes particularly challenging during large, extended demand surges, such as those caused by pandemics or natural disasters, due to the highly uncertain nature of the demand and potential for sudden spikes. Given the importance and difficulty of effective capacity management, it is crucial that hospitals have good decision-making processes in place. Capacity is typically managed manually by hospital administrators, often using select data and metrics to inform their decisions, but generally without prospective quantitative modeling and planning \citep{green2005}. There has been considerable research on developing mathematical models to aid capacity management decisions~\citep{mills2020, lacasa2020, nezamoddini2016, marquinez2021, sun2014, parker2024cm}; however, little work has been done toward implementing these models as part of a practical decision-making tool. Critically, while data and models can be valuable tools, capacity management decisions are ultimately made by hospital administrators, not by computational models, making effective communication of the insights from models as important as the model solutions themselves.

To address this problem, we develop a practical decision-support tool that translates the information from multiple capacity management models into a format that is informative to decision-makers. Models are used to predict demand volume, optimize decisions, and estimate important outcomes. The interface is an interactive dashboard that allows users to run these models in real-time with different parameters and forecast scenarios, and presents them with accessible visualizations of the results, along with key metrics. The dashboard enables users to explore and experiment with strategies, and helps them extract insights from the data and model outputs. A participatory design approach with hospital administrators ensured the tool provided sufficient breadth and granularity of information as well as actionable insights into potential decisions. The flexibility of the tool enables customization to specific goals, preferences, or other operational requirements of hospitals.

The dashboard was developed and implemented within a hospital system during the height of the COVID-19 pandemic in 2020 and 2021. During this period, there was an increased need in hospitals for capacity management tools due to unprecedented surges in demand and the drastic measures they necessitated. However, this need persists as hospitals continue to face significant surges owing to outbreaks of COVID-19 and other viruses, natural disasters, and climate-related events like extreme heatwaves. With the threat of pandemics, epidemics, and climate disasters growing, it is essential that hospitals have access to tools to help them respond to new situations. To facilitate implementation of similar tools in other hospital systems we have made the dashboard's code publicly available.\footnote{URL omitted during review process.}
% \footnote{\url{https://github.com/flixpar/HospitalCapacityManagementDashboard}}

Hospitals have many interventions they can use to manage capacity utilization, which generally fall into two categories: increasing capacity and reducing occupancy \citep{green2005, jack2009}. Hospitals have some ability to increase or redistribute capacity for specific services when needed by adjusting staffing, adding beds, shifting resources between services or units, and other strategies \citep{jack2009, fogerty2022, larsson2019, humphreys2022, mccabe2020, smith-daniels1988}. Hospitals can control occupancy to some extent by transferring or diverting patients, and postponing or canceling elective procedures, among a variety of other strategies that slow admissions or increase discharges \citep{mills2020, jack2009, mccabe2020, kelen2006, kim1999}. Each of these interventions has complex effects, costs, and benefits that must be taken into account by decision-makers in various roles within the hospital system \citep{green2005, li2003}. In this work, we target two high-level decisions made by hospital administrators: how much to increase or re-allocate capacity and how many patients to divert or transfer to other hospitals.
These specific interventions are essential for effective capacity management during surges, were frequently used during the COVID-19 pandemic, and would benefit from strategic planning.

This work aims to bridge the gap between existing operational hospital dashboards, which are limited in the types of information and insights they can provide, and hospital capacity management models, which are rarely used in practice due to the significant effort required to integrate them with real-time data and forecasts, as well as a lack of user-friendly interfaces for decision-makers.
Specifically, our goal is to improve hospital capacity management during surges by providing decision-makers with actionable information and insights from forecasting and optimization models in an accessible, real-time dashboard that enables interactive use of the models to investigate strategies and scenarios.

%% Related Work %%
\section{Related Works}
\label{sec:relatedwork}

Hospital capacity management encompasses a range of complex decisions related to patient flow, inter- and intra-hospital transfers, staffing, bed management, resource allocation, and more. Given the critical importance of these decisions in hospital systems, considerable attention has been devoted to methods that support hospital administrators, clinicians, and other stakeholders in making these decisions as effectively as possible.
Many studies have developed quantitative models to predict outcomes or optimize decisions. Other studies have created practical tools, often implemented as dashboards, to provide decision-makers with data and metrics to inform their decisions.
However, this study is one of the first to deeply integrate these strands of research.

\subsection{Decision-Support Models for Hospital Capacity Management}
\label{sec:relatedwork:models}

This study builds on the rich literature around quantitative data-driven models for hospital capacity management. Many angles of this problem have been explored, including demand forecasting \citep{klein2023, vollmer2021, weissman2020, kellner2023, qian2021}, staff scheduling \citep{miller1976, jaumard1998, aickelin2004, legrain2014}, resource allocation \citep{parker2020}, patient flow optimization \citep{bertsimas2024, day2010, zhang2020d, shi2016}, and demand smoothing \citep{helm2014, meng2015, chan2012, coffey2019, shi2021}.

Demand forecasting and smoothing are two key areas that lay the groundwork for effective capacity management. Forecasting methods, which predict demand for specific patient populations, typically employ machine learning models, statistical approaches, simulations, or combinations thereof \citep{klein2023, vollmer2021, kellner2023, qian2021, zhu2017, jones2008, weissman2020}. Complementing these efforts, demand smoothing techniques aim to reduce peak demand by strategically shifting patient volume across time. This can involve managing elective admissions, optimizing discharge timing, or a combination of both \citep{helm2014, shi2021, mills2020}. Within hospitals, patient flow optimization further contributes to capacity management by improving the efficiency of resource utilization \citep{bertsimas2024, day2010}.

Most relevant to our work are studies that model capacity allocation or inter-hospital patient transfer decisions during demand surges.
Capacity allocation involves determining how to distribute limited resources among different patient populations. Several studies have explored strategies for allocating beds between patient groups, or surgical capacity between elective and emergency care \citep{bekker2017, melman2021, ayvaz2010}.
For example, \citet{bekker2017} investigated four strategies for allocating beds between different patient groups to improve flexibility in adapting to normal demand fluctuations. \citet{melman2021} explored capacity allocation strategies through a simulation model examining the impact of canceling elective procedures and dividing operating room capacity between COVID-19 and elective surgeries. 
The utilization of inter-hospital patient transfers for capacity management has gained attention, particularly since the COVID-19 pandemic. Researchers have investigated how to optimize transfers to reduce strain on hospitals in various settings \citep{sun2014, michelson2021, marquinez2021, nezamoddini2016}. Some studies have focused specifically on load balancing between hospitals or regions during pandemics \citep{lacasa2020, dijkstra2023}. \citet{marquinez2021} and \citet{nezamoddini2016} develop Markov Decision Process and Mixed Integer Programming models, respectively, to reduce crowding and improve hospital utilization during non-surge periods. A notable contribution in this area is the work of \citet{parker2024cm}, which introduced a forecast model and decision optimization model for patient transfers and capacity allocation that incorporates many critical practical considerations.

While these methods show potential in simulations to significantly improve capacity management, few incorporate sufficient practical considerations to be applicable in real-world settings. Moreover, they are typically not integrated with interfaces that allow decision-makers to utilize them effectively. Our work aims to bridge this gap by adapting the model proposed by \citet{parker2024cm} to create a decision support tool that is both theoretically sound and practically implementable. We selected this model as the basis for our tool because it captures two types of decisions most relevant in our setting, includes pertinent practical constraints and objectives, and offers the flexibility to be readily adapted to our specific context. The details of our modifications and implementation are discussed further in \cref{sec:methods}.

\subsection{Decision-Support Dashboards for Hospital Capacity Management}
\label{sec:relatedwork:dashboards}

The development of decision-support dashboards for hospital capacity management has gained significant attention in recent years, as healthcare organizations seek to optimize resource allocation and improve operational efficiency. These tools have the potential to revolutionize decision-making processes by delivering actionable insights to stakeholders in an accessible, user-friendly format. While many dashboards are designed to support clinical decision-making \citep{dowding2015, park2022, badgeley2016}, there is a growing focus on developing dashboards to assist in management decisions, particularly in the context of hospital capacity planning and resource allocation \citep{rabiei2022, buttigieg2017}.
These dashboards largely aim to either improve situational awareness or evaluate process improvements.

Situational awareness is critical for real-time decision-making. For example, decision-makers managing resource use and patient flow must have reliable, up-to-date data on patients and resources.
Several dashboards have been developed to meet this need and inform decision makers in various contexts \citep{dixit2020, jawa2021, shahpori2013, gazivoda2022, kane2019, boudreault2023, franklin2017}.
\citet{dixit2020} detail the rapid development of visualization dashboards to support telehealth operations across a hospital system during the COVID-19 pandemic. These dashboards provided crucial real-time information on patient case load, wait times, and provider staffing to inform decisions regarding resource allocation and service scaling. This enhanced situational awareness and facilitated decision-making for healthcare system executives, telehealth leaders, and operational managers, demonstrating the effectiveness of user-centered design and iterative development in a crisis context.
Other dashboards, such as those developed by \citet{jawa2021} and \citet{shahpori2013}, specifically target ICU bed occupancy and ventilator utilization, which are critical resources during public health emergencies.
While these dashboards can improve situational awareness, they tend to provide information through a few simple metrics or visualizations, providing limited insights and assistance in the decision-making process.

Hospital performance dashboards have received significant attention as well \citep{buttigieg2017}. These dashboards tend to track key performance indicators (KPIs) over time to monitor trends and retrospectively evaluate the performance of interventions that have been implemented.
\citet{martinez2018} and \citet{franklin2017} developed dashboards that display KPIs related to patient flow and resource utilization, allowing hospital staff to quickly identify bottlenecks and make informed decisions.
\citet{weiner2015} describes the implementation of a data-driven dashboard which report a variety of KPIs including patient length of stay, emergency department throughput, diagnostic turnaround times, infection rates, and hand hygiene compliance. It enables hospital leadership, administrators, and staff, who use the dashboard to monitor and improve hospital performance across departments and units.

Some dashboards incorporate more advanced analytics and simulation capabilities to support capacity planning and resource allocation decisions.
For instance, \citet{weissman2020} develop a compartmental Monte Carlo simulation model for COVID-19 forecasting that predicts hospital-level admissions, census, and resource use. They integrated this model with a public online dashboard that allows users to modify the simulation parameters and see the predicted impacts in real-time.
\citet{toerper2018} created a web-based simulation tool that allows hospital administrators to estimate required surge capacity and evaluate the impact of different response strategies during mass casualty events. The tool allows users to interactively input a range of data, parameters, and potential decisions, and predicts the resulting available capacity. However, it is not able to optimize the decisions and produces limited outcome information.

Despite these advancements, most existing dashboards rely on reporting relatively simple metrics with standard visualizations, often failing to leverage the powerful decision-support forecasting, optimization, and planning models that have been developed for capacity management. Additionally, few dashboards have been specifically designed to address the unique challenges posed by pandemic surges and similar events, which require a higher level of adaptability and responsiveness in decision-making.

\subsection{Contributions}

Our work aims to bridge the gap between theoretical capacity management models and practical hospital dashboards. We developed a framework that integrates forecasting and optimization models with an interactive dashboard. The primary contributions of this work are:
\begin{enumerate}
    \item We adapted capacity management optimization models for practical real-time use. This involved modifying the models to optimize higher-level decisions and to allow for quick computation while maintaining the core functionality.
    \item We built an interactive dashboard that allows decision-makers to experiment with and use powerful prediction and modular optimization models without requiring technical expertise. Users can adjust model components and parameters, explore scenarios, and visualize the predicted outcomes of their decisions.
    \item We designed visualizations to effectively communicate key data and insights from the models. These visualizations balance the need for ease of use with the ability to perform detailed analysis, maximizing the utility of the tool across different levels of technical expertise.
    \item We integrated the dashboard with real-time hospital data systems and implemented it during the COVID-19 pandemic. This application provided insights into the challenges of deploying analytics tools in healthcare settings.
    \item We ensured the tool met the practical needs of its users by using a participatory design approach, involving hospital stakeholders throughout the development process.
\end{enumerate}

% These contributions demonstrate how advanced analytics can be made accessible to hospital administrators, potentially improving resource allocation during both routine operations and crises. The lessons from this implementation may inform future efforts to deploy similar systems in other healthcare settings.

% Our work aims to bridge the gap between the powerful theoretical models described in \cref{sec:relatedwork:models} and the hospital dashboards described in \cref{sec:relatedwork:dashboards}. To do so, we developed a framework that integrates forecasting and practical decision optimization models with an interactive, user-friendly dashboard. The primary contributions of this work are:
% \begin{enumerate}
% 	\item We adapted a set of capacity management optimization models for practical real-time use.
% 	\item We built an interactive dashboard to enable decision-makers to experiment with and use the models.
% 	\item We designed intuitive visualizations to effectively communicate the key data and insights.
% 	\item The framework was integrated with real-time data and implemented in practice during the COVID-19 pandemic.
% \end{enumerate}

%% Methods %%
\section{Data and Methodology}
\label{sec:methods}

\begin{figure}[htb]
    \centering
    \includegraphics[width=.85\textwidth, trim={11mm 3mm 7mm 6mm}, clip=true]{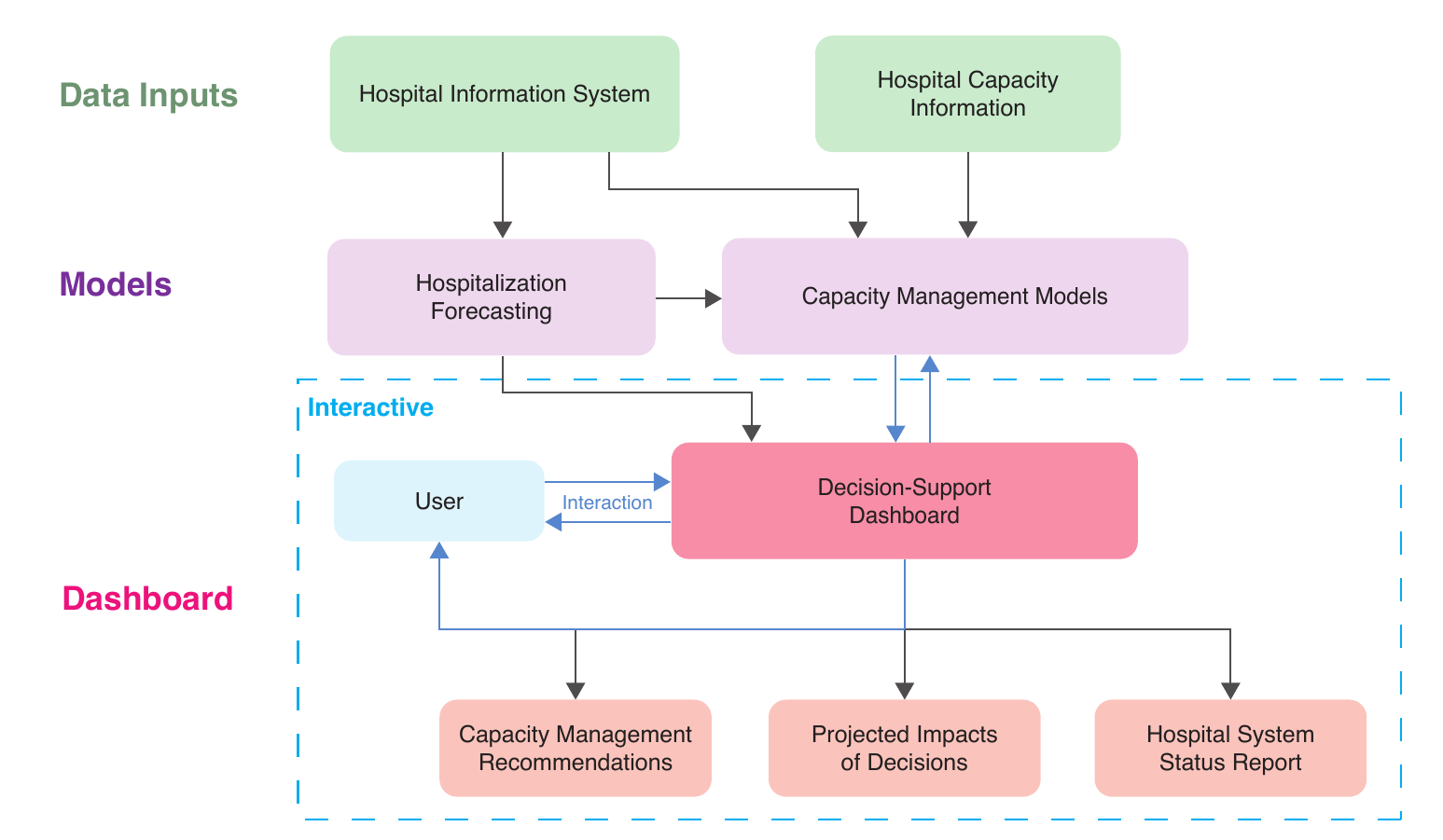}
    \caption{The interactive capacity management dashboard has three main components: ``Data Inputs'', ``Models'', and ``Dashboard''. The input data (green) is collected through EHR and HIS and fed into behind-the-scenes models shown in purple (the forecasts and optimization models). The user interacts with the tool through the Dashboard component (red) -- the focus of this work --and can investigate and examine the outcomes including recommended decisions and impact on operations (orange).}
    \label{fig:overview}
\end{figure}

The central focus of this work is the interactive decision-support dashboard for hospitals and healthcare systems. \Cref{fig:overview} provides an overview of the dashboard with the data and informatics components that are involved. First, data is collected from each hospital in the system (mainly through EHR and HIS), specifically admissions, occupancy, and capacity broken down by patient population and bed type. Second, this data informs a forecast model that predicts future demands. Third, the data is used to run models that determine optimal capacity management decisions. Finally, the output of each component feeds into the dashboard which users can interact with to help inform decision-making.

\subsection{Setting and Context}
\label{sec:methods:setting}

This study was performed in the Johns Hopkins Health System (JHHS) between August 2020 and May 2021. The health system is comprised of two urban academic hospitals and three suburban community hospitals, managing over 95,000 inpatient admissions, 337,000 emergency visits, and 931,000 primary and specialty care visits annually. We worked closely with the Johns Hopkins Hospital (JHH) Capacity Command Center (CCC) \citep{kane2019}, which is responsible for both real-time and strategic planning capacity management, and coordinates with other hospitals in the system. While the CCC uses state-of-the-art dashboards to track real-time and retrospective data, there are no existing tools that integrate interactive prospective modeling to aid in planning.

\subsection{Participatory Design Process}
\label{sec:methods:participatory}

\begin{figure}[htb]
    \centering
    \includegraphics[width=.8\textwidth, trim={11mm 3mm 8mm 8mm}, clip=true]{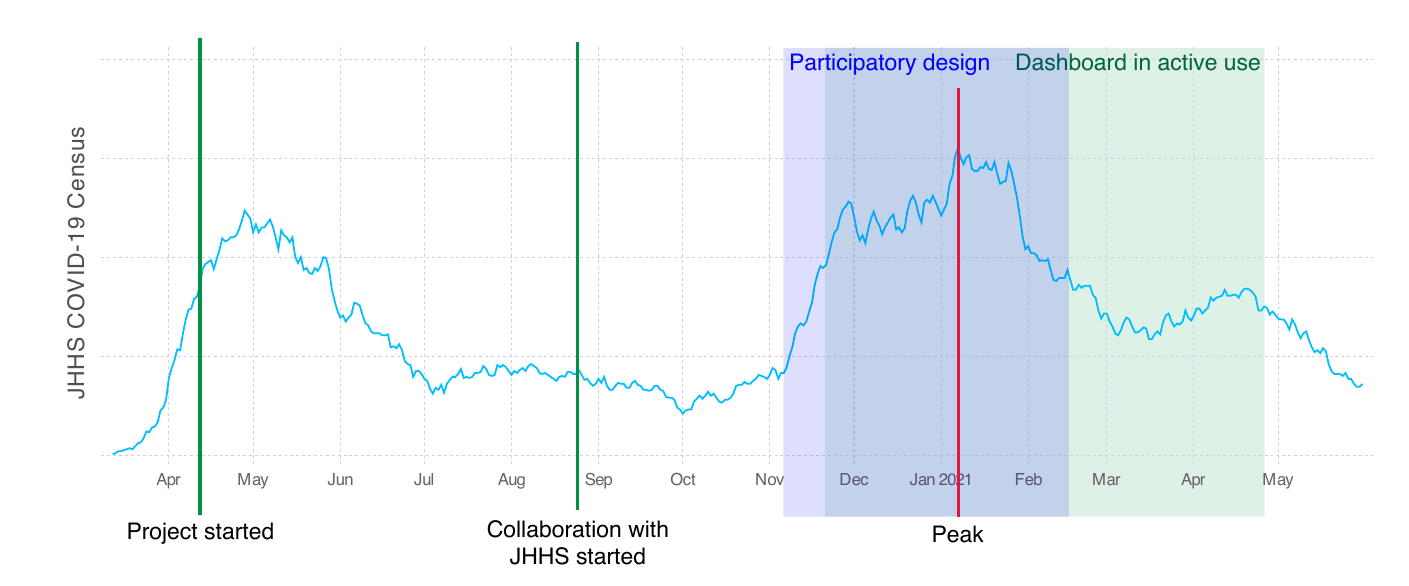}
    \caption{Overview of the project timeline with key phases and events. The timeline is overlayed over the total JHHS COVID-19 patient occupancy for context. An early version of the model was developed in April 2020. Initial work towards a dashboard catered to JHHS began in August 2020, with active development and use of the dashboard beginning in November 2020.}
    \label{fig:project-timeline}
\end{figure}

The models and dashboard interface that make up our tool were developed through a participatory design process \citep{schuler1993participatory} between the systems engineering team, end-users of the dashboard within the modeling working group of the CCC, and stakeholders throughout JHHS. This process spanned from November 2020 to February 2021 (\cref{fig:project-timeline}) when there was increased operational need for tools to help inform capacity management decisions due to the COVID-19 pandemic. The engineering team was responsible for designing and implementing an initial version of the dashboard, including the interface and model, and then incorporating feedback into future iterations of the dashboard. The researchers in the modeling group were the primary end-users of the dashboard and used it to investigate different demand scenarios and capacity management strategies. Finally, the stakeholders were informed of projections and recommendations by members of the modeling working group in weekly meetings. The design of the dashboard was iterated repeatedly to ensure that the dashboard was easy to use and valuable while taking into account the operational concerns of the end-users and stakeholders. This process improved many aspects of the tool from the types of decisions that were considered, to the details of the model, and the design of the figures displayed on the interface. These improvements will be reviewed in the discussion section.

Through our participatory design process, two types of capacity decisions stood out: determining the appropriate surge capacity level and the minimum number of operational patient transfers. Identifying the capacity level was the primary way JHHS hospitals adapted to COVID-19 surges. The JHHS had identified six surge levels, but the criteria for transitioning between them were not set and decisions were made in real-time. Higher surge capacity levels, while necessary, were undesirable as converting capacity was an expensive and time-consuming process that took resources away from other patients. That posed a challenge because the size and timing of surges were uncertain. Therefore, it was critical to optimize the timing of each surge level increase or decrease. In parallel and as another response measure, the CCC was also interested in using patient diversions (i.e., redirecting incoming patients to another hospital) or transfers (i.e., redirecting admitted patients to other hospitals with a higher level of care) to protect the smaller hospitals in the system from getting overrun. Neither of these interventions had been deployed for operations reasons previously, therefore, a decision-support tool was especially valuable in helping JHHS develop strategies to implement them. The implementation details of how to reach the desired capacity level or which specific patients to divert are up to managers and clinicians.

\subsection{Data Inputs and Processing}
\label{sec:methods:data}

The primary inputs to the decision-support models are the number of daily admissions for each hospital and the capacity levels of each hospital (see \cref{tab:inputs} for more details). The admissions (required for the prospective analysis) are obtained from Hospital Information Systems (HIS) and divided into relevant patient groups, e.g., ICU patients, mechanical ventilator patients, or confirmed COVID-19 patients. The tool does not use patient-level data or other sensitive information that must be protected, so these inputs are generally reliable and easy to obtain. Prospective analysis is done using forecasts of admissions. In our implementation, we employ internal forecasts from JHHS to predict COVID-19 hospitalizations at the facility level for up to three months into the future. We use a two-week short-term forecast for maximal accuracy in planning upcoming decisions, and a long-term multi-scenario forecast for planning beyond that.

\begin{table}[ht]
    \centering
    \resizebox{\linewidth}{!}{%
    \begin{tabular}{>{\hspace{0pt}}m{0.2\linewidth}|>{\hspace{0pt}}m{0.22\linewidth}>{\hspace{0pt}}m{0.13\linewidth}>{\hspace{0pt}}m{0.5\linewidth}}
        \toprule
        \textbf{Data Type}             & \textbf{Source}         & \textbf{Resolution} & \textbf{Uses}                                        \\ 
        \hline %midrule
        Patient arrivals               & HIS / forecast          & Hospital, day       & Dashboard, optimization model, parameter estimation  \\
        Patient census                 & HIS / forecast          & Hospital, day       & Dashboard, parameter estimation                      \\
        Hospital capacity              & Estimate from hospitals & Hospital            & Dashboard, optimization model                        \\
        Patient length of stay & Estimate from EHR       & Hospital            & Optimization model    \\
        \bottomrule
    \end{tabular}}
    \caption{Summary of each input data type used in our decision-support tool.}
    \label{tab:inputs}
\end{table}

The available capacity can be difficult to determine, particularly when focusing on specific patient populations, such as COVID-19 patients. Hospitals may have a fixed number of physical beds, but true capacity depends on staffing and resources, and to some extent can be re-allocated between departments, services, and patient populations when needed, especially during large surges. In preparation for COVID-19 outbreaks, JHHS adopted a set of ``surge levels'', along with a list of units that would be used for COVID-19 patients at each hospital in each surge level. In addition to the baseline level, there were five possible surge levels: ``Ramp-Up'', ``Surge'', ``Surge+'', ``Maximum'', and ``Crisis''. Since the underlying models and dashboard framework are flexible to consider any arbitrary units of capacity, we updated them to reflect the six capacity levels in JHHS and to select the most appropriate level for each patient type and time point.

\subsection{Capacity Management Models}
\label{sec:methods:model}

The second component of this tool is the behind-the-scenes technical framework to forecast surge demand and optimize decisions.
We utilized an internal forecast of COVID-19 admissions at each hospital in JHHS that was developed by a team within the system during the early stages of the pandemic to predict incoming surge demand. It used a large agent-based simulation that took into account many factors including mobility levels, lockdown orders, school closures, vaccination rates, and more. Projections were made for a range of scenarios by adjusting assumptions about how mobility, policy, and vaccination rates would evolve.

Hospitalization forecasting models can predict hospitalization trends but do not directly quantify the downstream effects of those incoming patients on a hospital system. We complement the forecast model with data-driven capacity management models to estimate those effects, mitigate them, and answer questions for decision-makers.
The results from the models estimate an array of outcomes including hospital occupancy over time, expected number of admissions and discharges, required daily surge capacity levels, and whether a hospital has enough total surge capacity, among others. The models are not only used to find the optimal decisions concerning these estimated outcomes, but also to answer additional questions. For example, the impact of transfers on surge capacity levels, by varying the number of transfers and analyzing the corresponding outcomes. Additional analytics are also provided on the data and outcomes, including admission targets for each hospital on each day which signals the approximate number of patients a hospital can admit daily before they need to start diverting patients or risk an overflow.

Our model is based on the robust mixed-integer linear optimization model introduced in \citet{parker2024cm}, which simultaneously selects the optimal set of ``units'' or rooms to allocate to the given patient population and number of patients to transfer between each pair of hospitals over time.
We adapted this model to the particular setting we target in this work, focusing on the needs of JHHS during peaks in the COVID-19 pandemic. We selected operational constraints, objectives, and parameters through the participatory design process, modifying the model iteratively to suit the needs of the end-users of the dashboard while ensuring a modular implementation for easy end-user customization. More details about the specific operational constraints and parameters used are given in \cref{tab:params}.
See \cref{appendix:model} for the full formulation of this model.

We also developed a simplified model to fit the real-time setting of the dashboard.
The simplified model eliminates low-level room or unit opening and conversion decisions, which were typically made manually, and relaxes integrality constraints on the number of transfers. In practice, the high-level decisions about surge level and transfers were largely consistent between models, so users could experiment with the faster model, then refine plans using the complete model.
Users were given the option of running the complete MIP model, which typically took 30-60 seconds, or the simplified model, which completed in 1-2 seconds, making it far easier to experiment and iterate on plans. The option was given to users through the ``Model Complexity'' select displayed in \cref{fig:dashboard-params} and explained in \cref{tab:params}.
The formulation of the simplified model (without the practical constraints and objectives) is presented in \crefrange{eq:obj}{eq:end}. The notation used is described in \cref{tab:notation}.

\begin{table}[tb]
	\centering\footnotesize
	\begin{tabular}{p{1.4cm}p{3.6cm}p{10.8cm}}
		\toprule
		Notation			& Name		& Description		\\
		\midrule
		\( \T \)		& Time steps		& Set of all time steps (days)		\\
		\( \H \)		& Hospitals		& Set of hospitals		\\
		\( \mathcal{L} \)		& Surge Levels		& Set of surge levels		\\
		\( o_{h,t} \)		& Census		& Census of hospital $h$ at time $t$		\\
		\( b_{h,l} \)		& Capacity per surge level		& Number of staffed beds at hospital $h$ in surge level $l$		\\
		\( i_{h,t} \)		& Incoming demand		& Number of patients arriving at hospital $h$ during time period $t$		\\
		\( L_h \)		& Length of stay		& Distribution over length of stay for patients at hospital $h$		\\
		\( c_{h,t} \)		& Allocated capacity		& Capacity allocated at hospital $h$ during time period $t$		\\
		\( u_{h,t,l} \)		& Surge level		& Binary variable indicating whether hospital $h$ is at surge level $l$ during time period $t$		\\
		\( s_{h,g,t} \)		& Patient transfers		& The number of patients to transfer/divert from hospital $h$ to hospital $g$ during time period $t$		\\
		\( w^{(1)}, w^{(2)} \)		& Objective function weights		& Costs or weights for the optimization objective function		\\
		\bottomrule
	\end{tabular}
	\caption{Mathematical notation used in the optimization model.}
	\label{tab:notation}
\end{table}

\begin{subequations}
\begin{align}\small
	\text{min}\quad & \sum_{h \in \H} \sum_{t \in \T} \sum_{l \in \mathcal{L}} w^{(1)}_{h,l} u_{h,t,l} + \sum_{h \in \H} \sum_{g \in \H} \sum_{t \in \T} w^{(2)}_{h,g} s_{h,g,t} & \label{eq:obj} \\
	\text{s.t.}\quad &c_{h,t} \geq o_{h,t} &\forall h \in \H, t \in \T	\label{eq:cons:no-shortage} \\
	&c_{h,t} = \sum_{l \in \mathcal{L}} u_{h,t,l} b_{h,l} &\forall h \in \H, t \in \T	\label{eq:exp:capacity} \\
	&\{ u_{h,t,l} \}_{l \in \mathcal{L}} \in SOS1 &\forall h \in \H, t \in \T \label{eq:cons:sos} \\
	&o_{h,t} = \sum_{t^\prime = 1}^{t} \left( a_{h,t^\prime} - d_{h,t^\prime} \right)		&\forall h \in \H, t \in \T		\label{eq:exp:census} \\
	&a_{h,t} = i_{h,t} + \sum_{g \in \H} \left( s_{g,h,t} - s_{h,g,t} \right)			&\forall h \in \H, t \in \T		\label{eq:exp:admissions} \\
	&d_{h,t} = \sum_{t^\prime = 1}^{t} \left( P(L_h = t - t^\prime) a_{h,t^\prime} \right)		&\forall h \in \H, t \in \T	\label{eq:exp:discharges} \\
	&u_{h,t,l} \in \set{0,1}			&\forall h \in \H, t \in \T, l \in \mathcal{L} \\
	&s_{h,g,t} \in \mathbb{R}_{\geq 0} &\forall h,g \in \H, t \in \T \label{eq:end}
\end{align}
\end{subequations}

The decision variables in the simplified model are the surge level of each hospital ($h$) during each day ($t$), $u_{h,t,l}$, and the number of transfers between each pair of hospitals ($h,g$) during each day, $s_{h,g,t}$.
The objective (\cref{eq:obj}) is simply to minimize the cost of these decisions.
The model constrains capacity at each hospital to be at least as much as the estimated census (\cref{eq:cons:no-shortage}) so there is enough capacity for all patients, where capacity is determined from the surge level (\cref{eq:exp:capacity}) and census is determined from past admissions and discharges (\cref{eq:exp:census}).
The model uses a type 1 special ordered set (SOS1) constraint (\cref{eq:cons:sos}) to ensure each hospital is at a single surge level at each time point. SOS1 constraints can be reformulated as MIP constraints, but current MIP solvers also have specialized methods for efficiently incorporating these constraints.
All of the surge level, capacity, census, admissions, and other information used on the dashboard for metrics and visualizations can be computed using the expressions in this model.

\subsection{Decision-Support Dashboard}
\label{sec:methods:dashboard}

The main component of our decision-support tool is the dashboard interface that users interact with to investigate the data, projections, and recommendations of the models across a range of scenarios and parameters. The dashboard consists of three main pages: the decision recommendation tool, a data and forecast exploration page, and an overall status report.

\subsubsection{Decision Recommendation Tool}

\begin{figure}[htb]
    \centering
    \includegraphics[width=.78\textwidth, trim={12mm 4mm 7mm 4mm}, clip=true]{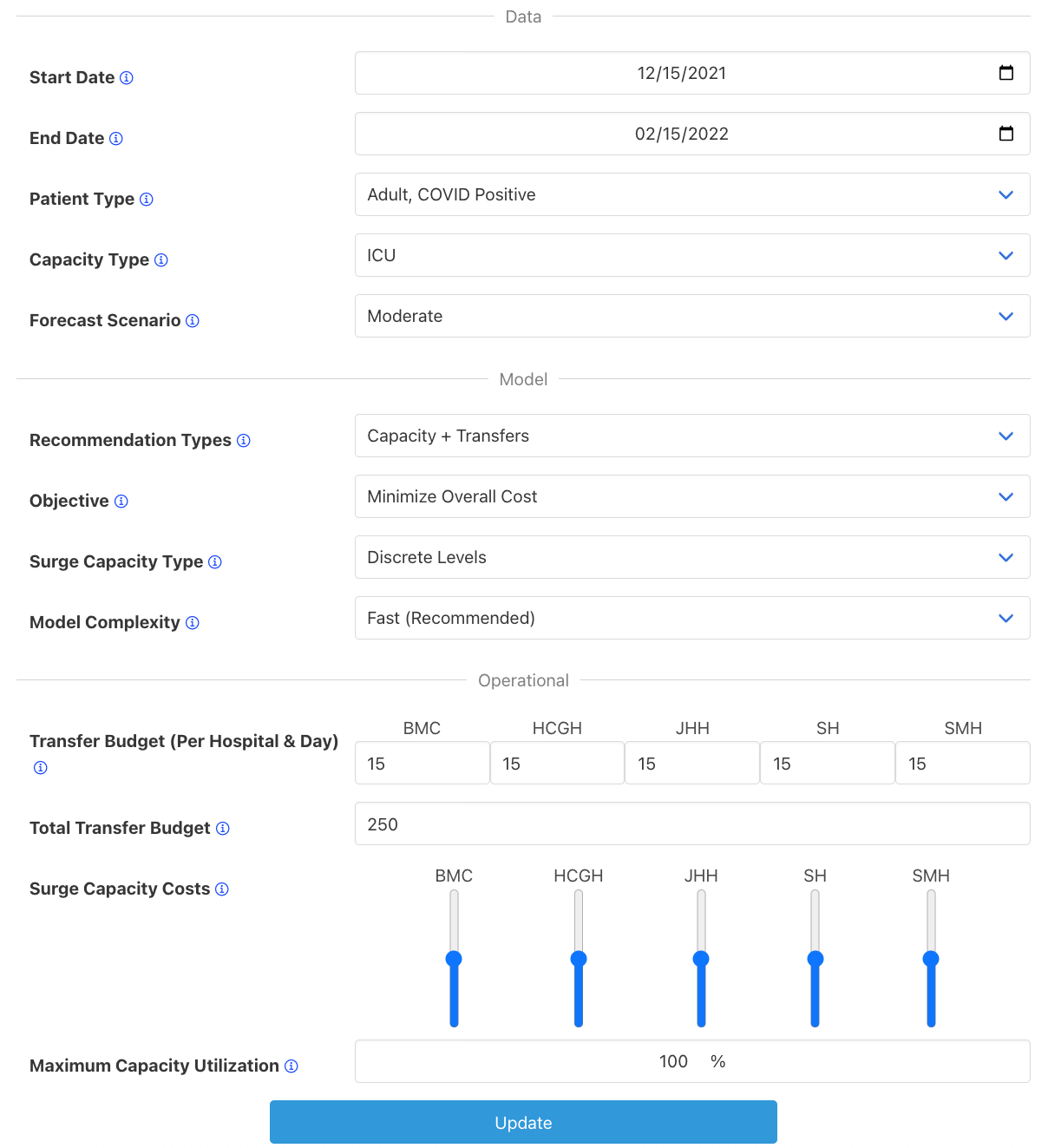}
    \caption{User interface to interactively customize the modular optimization model and data inputs. The description and possible values for each option are listed in \cref{tab:params}. When a user presses ``Update'', the new customized optimization model runs on the backend and all visualizations are updated with the new results.}
    \label{fig:dashboard-params}
\end{figure}

The decision recommendation page enables users to analyze the projected impact of capacity management decisions on the future state of the hospital system. It computes the optimal decisions under the constraints and preferences that users specify, and estimates how these decisions will affect outcomes and performance metrics.
Users can control 20 parameters that govern the data, forecast, and decision optimization process.
\Cref{fig:dashboard-params} presents the interface that users interact with the modular optimization to select parameters, objectives, and constraints based on their needs. The customizable inputs are grouped into those that relate to the data, those that select the model formulation, and those that adjust operational model parameters.
The interface uses a combination of select dropdown menus, date selectors, numerical inputs, and sliders to make parameter selection easy and intuitive.
For the operational model parameter selections, numerical inputs allow users to precisely control options that are interpretable and have absolute units, whereas the sliders enable users to easily set relative values without forcing them to pick precise numerical values.
The blue ``i'' symbols pop up descriptions of each option when users hover their cursor over the symbols to provide context.
Full details on each parameter are listed in \cref{tab:params}.

When a user changes the options and presses ``Update'', the customized optimization model runs on the backend and returns results. Once the model runs, the dashboard reports key metrics in three tables and helps the user interpret the results with ten visualizations, which are displayed and discussed in \cref{sec:methods:vis}. At any point, a user can re-run the model with updated parameters to investigate a different scenario or tweak the recommended decisions.

The parameters that users control include forecast scenarios, intervention types, optimization objectives, and the relative burden of surge capacity at each hospital. This control enables users to employ the tool in various ways to fit a range of use-cases. One of the primary uses of this decision-support tool has been for scenario analysis. The interactive and real-time nature of the tool enables users to test different forecast scenarios (e.g., a pessimistic forecast) and model parameters to explore how the selection changes the recommended decisions and the projected outcomes. As the number of patient arrivals cannot be known with certainty, planning for different possibilities is essential in ensuring that the hospital system can respond well to any scenario. Allowing the user to have a great deal of control over the model configuration ensures operational requirements and priorities are captured while guaranteeing the optimality of the recommended decisions.

\begin{table}[ht]
\centering
\footnotesize
\renewcommand{\arraystretch}{1}
\centerline{
\begin{tabular}{>{\RaggedRight\arraybackslash}p{0.15\textwidth}p{0.45\textwidth}>{\RaggedRight\arraybackslash}p{0.3\textwidth}}
\toprule
\textbf{User Input} & \textbf{Explanation} & \textbf{Possible Values} \\ 
\hline
Start Date & Start date for the analysis & Between March 1, 2020 and 2 months from today’s date \\ \hline
End Date & End date for the analysis & Between April 1, 2020 and 3 months from today’s date \\ \hline
Patient Population & Patient population to analyze & All, adult, pediatric, COVID-positive, adult COVID-positive, pediatric COVID-positive \\ \hline
Capacity Type & Type of capacity to focus on & Total, ICU beds, general beds \\ \hline
Forecast Scenario & Level of optimism/pessimism to assume when forecasting hospitalizations & Optimistic, moderate, pessimistic \\ \hline
Recommendation Type & Decision types that the model will try to optimize & Capacity + transfers, capacity only, transfers only, none \\ \hline
Model Objective & Determines whether the model tries to minimize the total burden of implementing the decisions, minimize the surge capacity required given the limits on transfers, or maximize fairness by distributing load evenly & Minimize overall costs/burden, minimize surge capacity, balance load evenly \\ \hline
Surge Capacity Type & Determines whether surge capacity must be determined by the surge level (discrete levels) or whether it can be adjusted at a bed level & Discrete levels, continuous \\ \hline
Model Complexity & Determines whether to solve the complete model or the faster simplified model & Fast, complete \\ \hline
Transfer Budget (per hospital-day) & Maximum number of incoming and outgoing transfers allowed for each hospital, during each day & $\geq 0$ for each hospital \\ \hline
Total Transfer Budget & Maximum number of transfers allowed during each day across the entire system & $\geq 0$ \\ \hline
Relative Surge Capacity Costs & Burden of adding surge capacity at each hospital relative to the others & 0 – 1 for each hospital \\ \hline
Maximum Capacity Utilization & Percentage of the total capacity each hospital can use in practice before running out of effective capacity & 0\% – 100\% \\
\bottomrule
\end{tabular}}
\caption{The users can modify these parameters of the modular optimization model through the decision-support page of the dashboard, allowing them to interactively plan for multiple scenarios and operational restrictions while controlling the recommendations. The parameter explanations were shown to users with pop-ups that appeared when hovering next to the parameter names in \cref{fig:dashboard-params}.}
\label{tab:params}
\end{table}

\subsubsection{Visualizations}
\label{sec:methods:vis}

The visualizations on the decision-support dashboard are designed to provide users with a clear, comprehensive view of the current and predicted state of the hospitals, as well as the impact of the recommended actions. Unlike standard, off-the-shelf visualizations, these figures were carefully crafted to present information in the most intuitive way possible while still providing sufficient detail to support operational decision-making. By presenting the information in a visual format tailored to the specific needs of the users, the dashboard enables individuals with different levels of technical expertise to quickly grasp the situation and make informed decisions. The visualizations were designed through an iterative process in close collaboration with end-users to ensure they met their unique requirements and were easy to interpret.

\begin{figure}[htb]
    \centering
    \begin{subfigure}[b]{0.75\linewidth}
       \includegraphics[width=\linewidth, height=0.45\textwidth, trim={0mm 1mm 1mm 0mm}, clip=true]{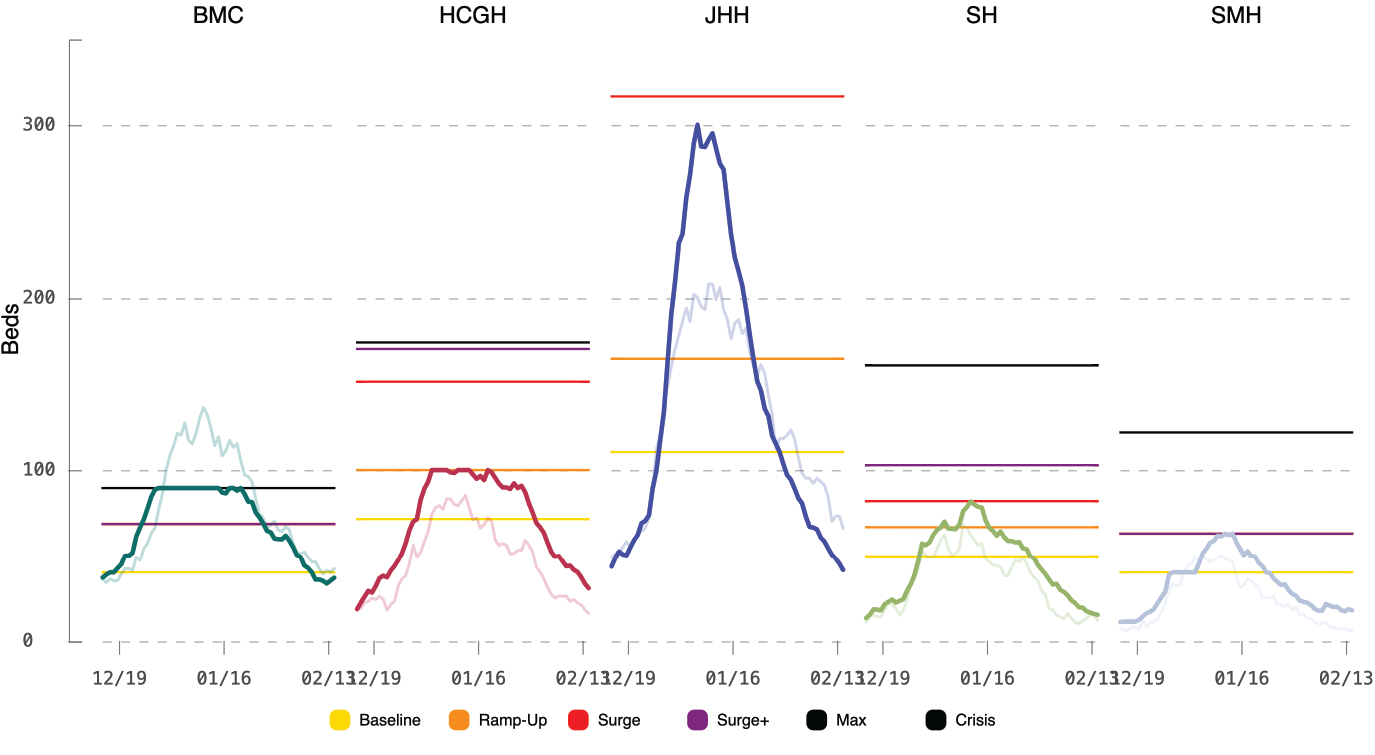}
        \caption{}
        \label{fig:visualizations:census:a}
    \end{subfigure}
    % \hspace{0.25em}
    \begin{subfigure}[b]{0.75\linewidth}
       \includegraphics[width=\linewidth, height=0.42\textwidth, trim={0mm 0mm 1mm 0mm}, clip=true]{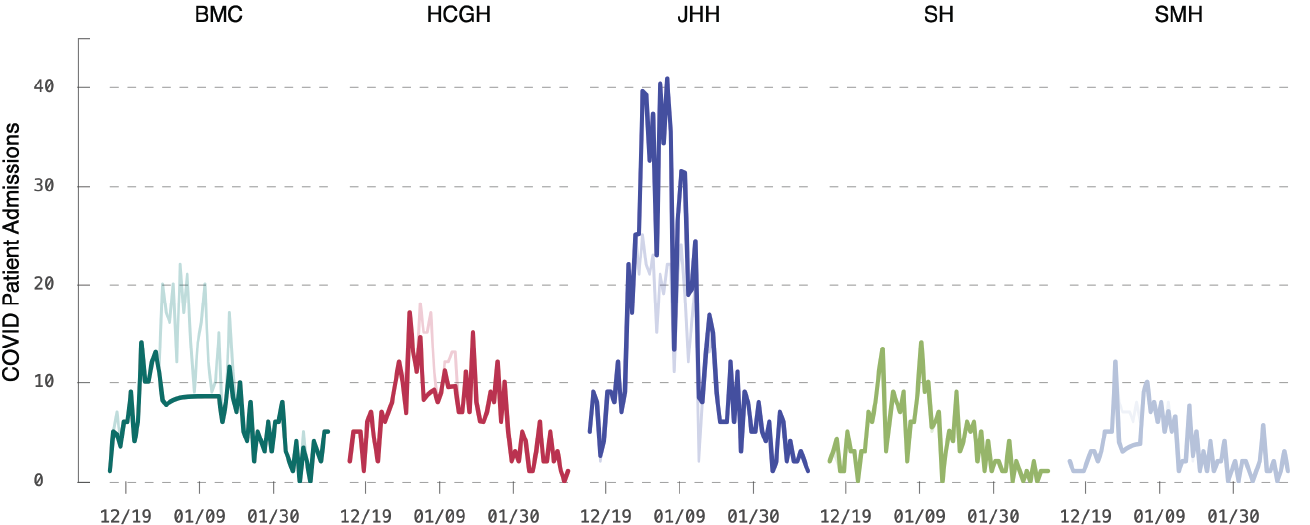}
        \caption{}
        \label{fig:visualizations:census:b}
    \end{subfigure}
    \caption{Overview of hospital census and admissions with and without transfers. Figures are screenshots of the visualizations taken from the dashboard. (a) Shows the projected COVID-19 census over time for each hospital, both with (dark curves) and without (light curves) patient transfers/diversions for each hospital, along with capacity levels. (b) Displays the COVID-19 patient admissions over time, with and without the recommended transfers (dark and light curves, respectively).}
    \label{fig:visualizations:census}
\end{figure}

\begin{figure}[htb]
    \centering
    \begin{subfigure}[b]{0.875\linewidth}
       \includegraphics[width=\linewidth, trim={1mm 2mm 1mm 1mm}, clip=true]{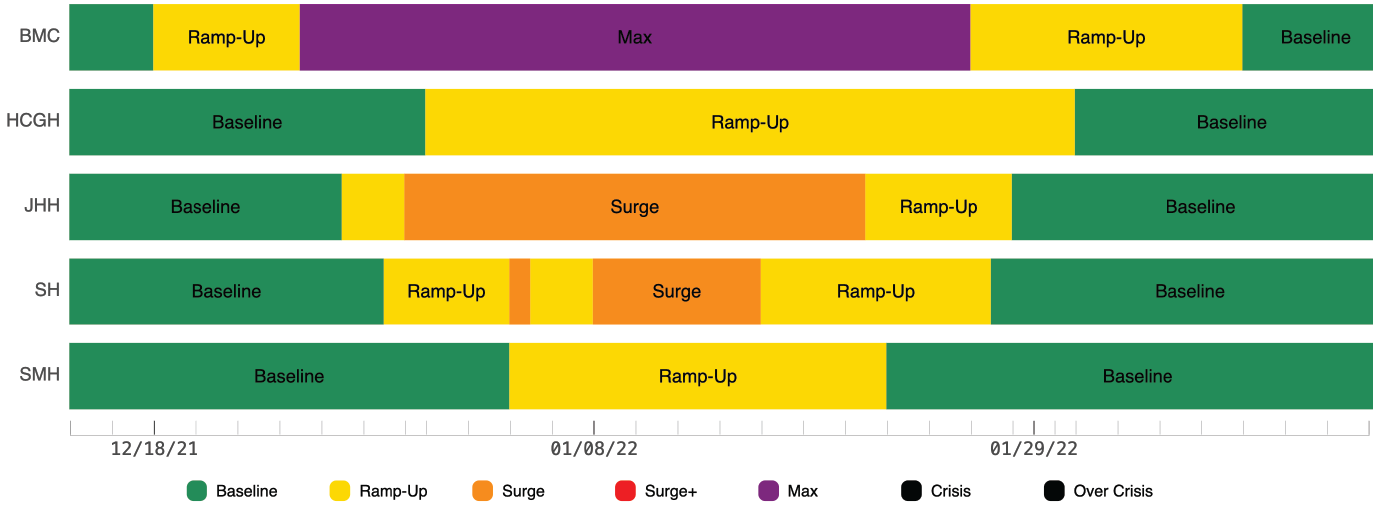}
        \caption{}
        \label{fig:visualizations:capacity:a}
    \end{subfigure}
    % \hspace{0.25em}

    \begin{subfigure}[b]{0.75\linewidth}
       \includegraphics[width=\linewidth, trim={3mm 17mm 7mm 10mm}, clip=true]{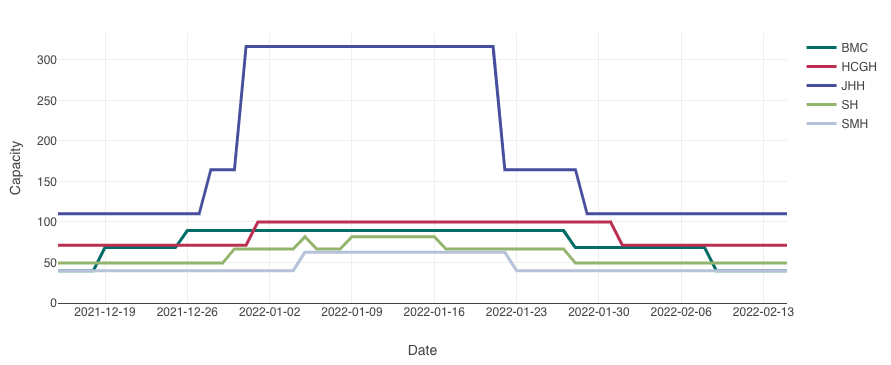}
        \caption{}
        \label{fig:visualizations:capacity:b}
    \end{subfigure}
    \caption{Screenshots of the visualizations for recommended hospital capacity and surge levels over time. (a) Highlights the surge level required for each hospital over time in a simplified way, which allows the hospitals to plan their surge capacity opening strategies. (b) Shows the required capacity for each hospital in more detail.}
    \label{fig:visualizations:capacity}
\end{figure}

\Cref{fig:visualizations:census} presents screenshots of the visualizations for high-level overviews of predicted patient census and admissions at each hospital. These figures allow users to quickly assess the projected state of each hospital in the system over the coming weeks, and how the census and admissions would be impacted by the recommended transfers. This allows hospitals to know what to plan for and to evaluate whether the recommended transfers would be sufficient to alleviate strain on hospitals.
These were chosen to help users understand the current and predicted occupancy of each hospital, and how the recommended transfers would affect these numbers. The key design decision here was to show the census and admissions both with and without the recommended transfers on the same plot. This allows users to easily compare the two scenarios and understand the impact of the transfers.
Users can use these visualizations in conjunction with the capacity visualizations (\cref{fig:visualizations:capacity}) to identify when a hospital is likely to reach capacity. Hover functionality, allowing users to see exact values, was added to provide additional detail when needed.

The capacity visualizations (\cref{fig:visualizations:capacity}) were designed to help hospitals plan their surge capacity strategies at different levels of detail. \Cref{fig:visualizations:capacity:a} provides a simplified, high-level view of when each hospital needs to increase its surge capacity. This was designed to give users a quick overview of the situation without overwhelming them with details. \Cref{fig:visualizations:capacity:b}, on the other hand, provides a more detailed view of the required capacity for each hospital over time. This view was included based on user feedback indicating that this level of detail was necessary for operational planning. The zooming functionality was added to allow users to focus on specific time periods of interest.

\begin{figure}[htb]
    \centering
    \begin{subfigure}[b]{0.49\linewidth}
       \includegraphics[width=\linewidth]{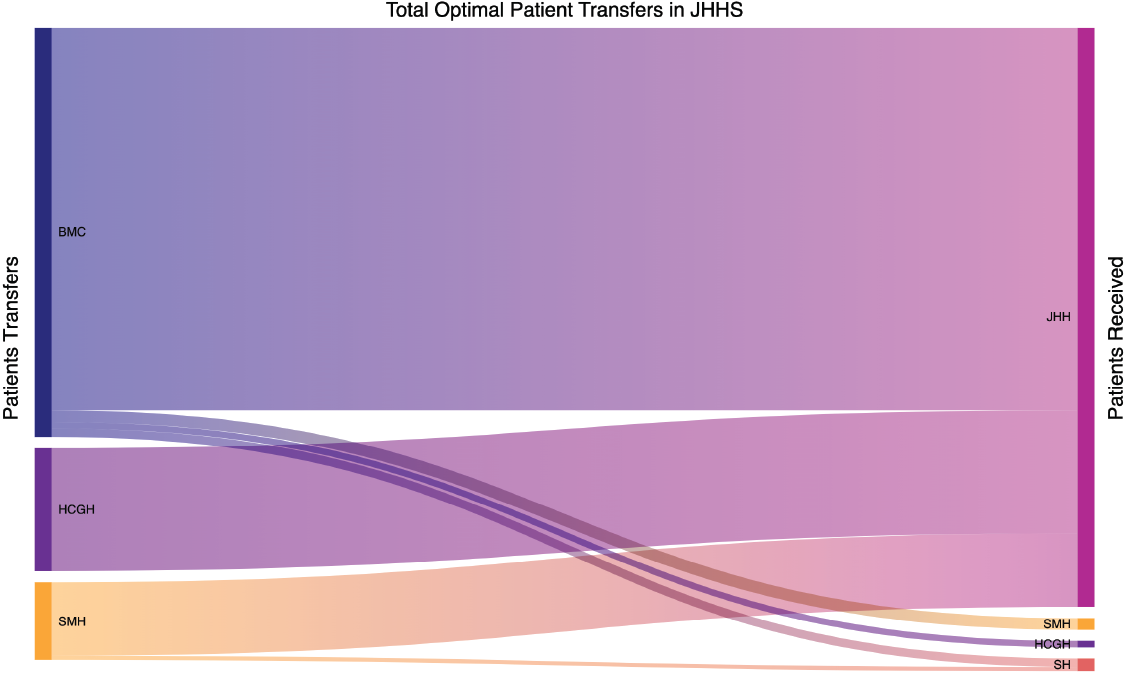}
        \caption{}
        \label{fig:visualizations:transfers:a}
    \end{subfigure}
    \hspace{0.25em}
\begin{subfigure}[b]{0.49\linewidth}
       \includegraphics[width=\linewidth, trim={1mm 0mm .5mm 0mm}, clip]{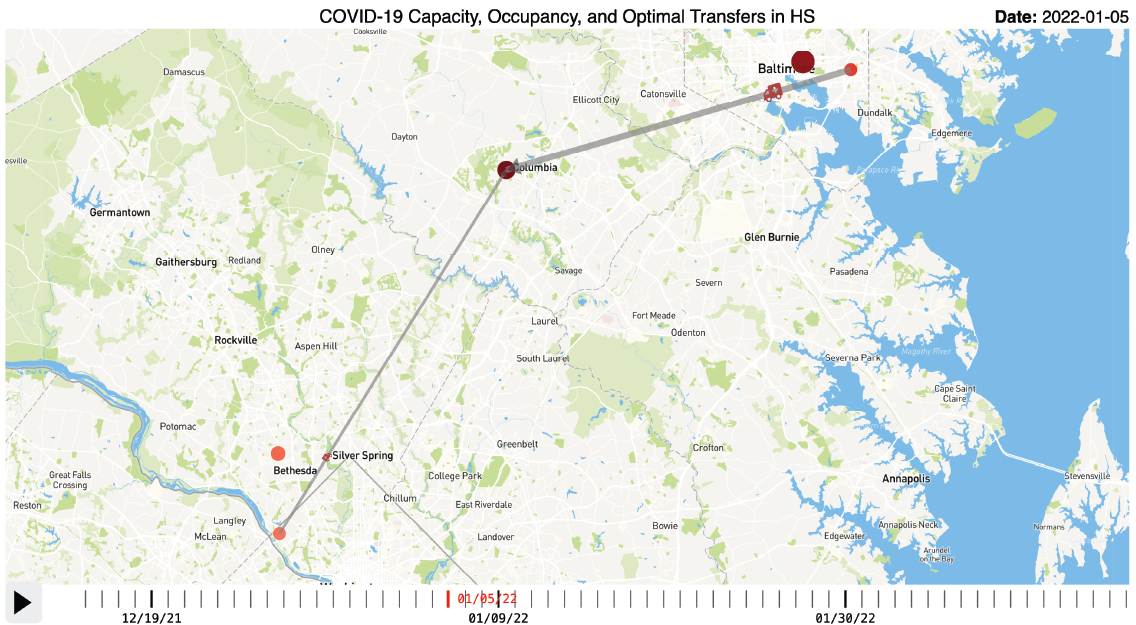}
        \caption{}
        \label{fig:visualizations:transfers:c}
    \end{subfigure}
    \begin{subfigure}[b]{0.49\linewidth}
       \includegraphics[width=\linewidth, height=.6\linewidth, trim={.5mm 1.5mm 1mm 1mm}, clip]{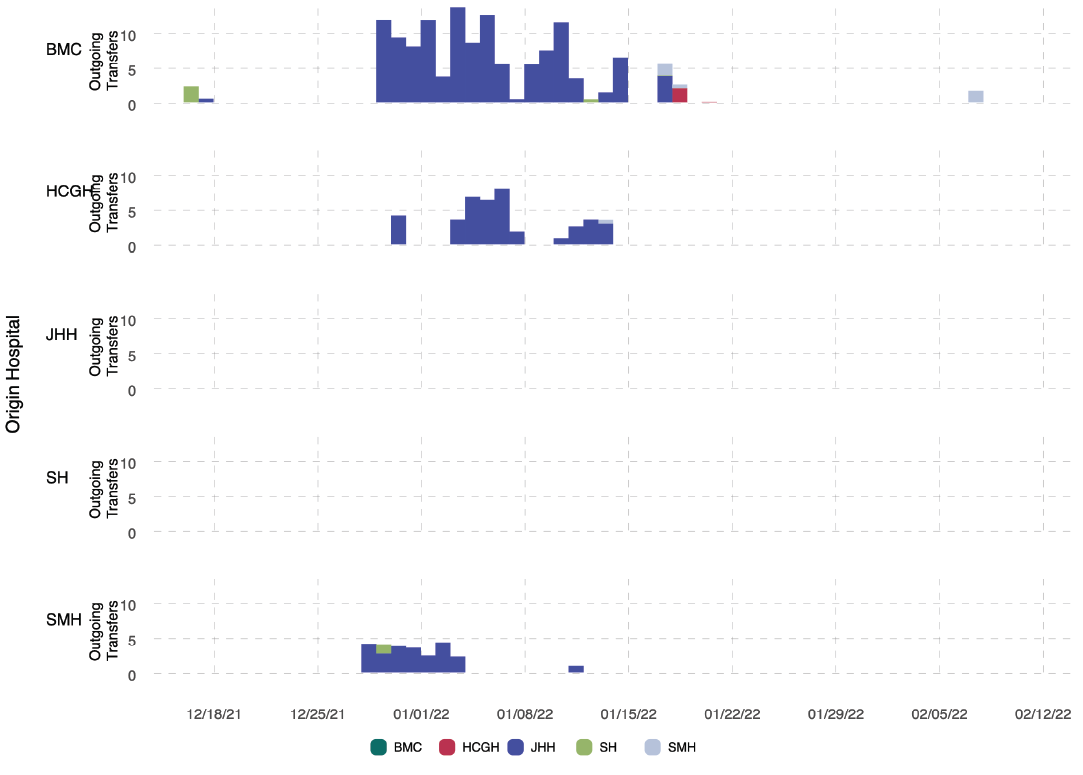}
        \caption{}
        \label{fig:visualizations:transfers:b}
    \end{subfigure}

    \caption{Screenshots of the visualizations for recommended patient transfers. (a) Provides an overview of the overall flow of recommended transfers. (b) Is an animated map that visualizes the recommended diversions/transfers the user selected over time. (c) Breaks down exactly how many patients should be transferred/diverted between each pair of hospitals during each day.}
    \label{fig:visualizations:transfers}
\end{figure}

\Cref{fig:visualizations:transfers} focuses on the recommended patient transfers. Several non-standard visualizations were used here to effectively communicate different aspects of the transfer recommendations. \Cref{fig:visualizations:transfers:a} uses a Sankey diagram to show the overall flow of transfers between hospitals. This visualization was chosen because it effectively communicates the volume of transfers between each hospital pair in a way that is immediately understandable, even for users who are not familiar with this type of diagram. \cref{fig:visualizations:transfers:c} uses an animated map to visualize the transfers geographically over time. This unique visualization was specifically designed for this dashboard to help users understand the spatial and temporal distribution of the transfers, which is important for identifying potential bottlenecks in the transfer process. Finally, \Cref{fig:visualizations:transfers:b} uses a heatmap to show the specific number of patients to be transferred between each hospital pair on each day. This provides a more detailed view of the transfer recommendations over time.

\begin{figure}[htb]
    \centering
    \begin{subfigure}[b]{0.55\linewidth}
       \includegraphics[width=\linewidth]{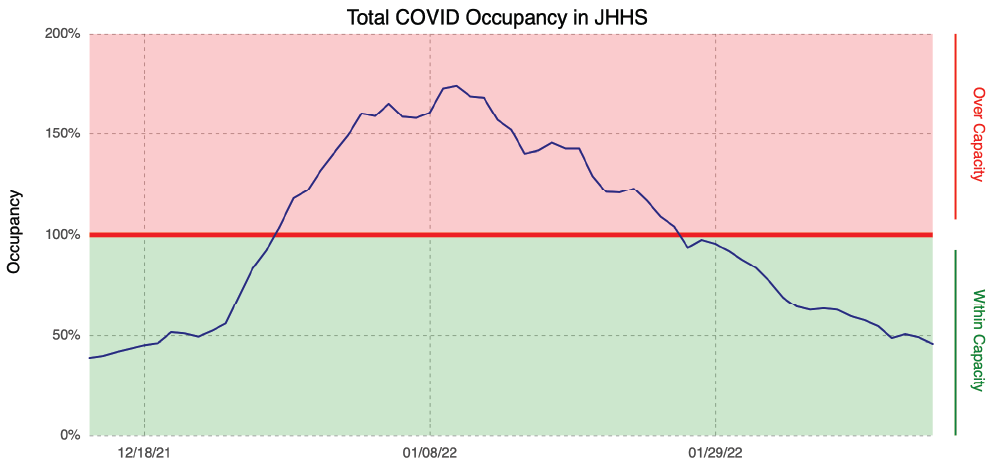}
        \caption{}
        \label{fig:visualizations:load:a}
    \end{subfigure}

    \begin{subfigure}[b]{\linewidth}
       \includegraphics[width=\linewidth]{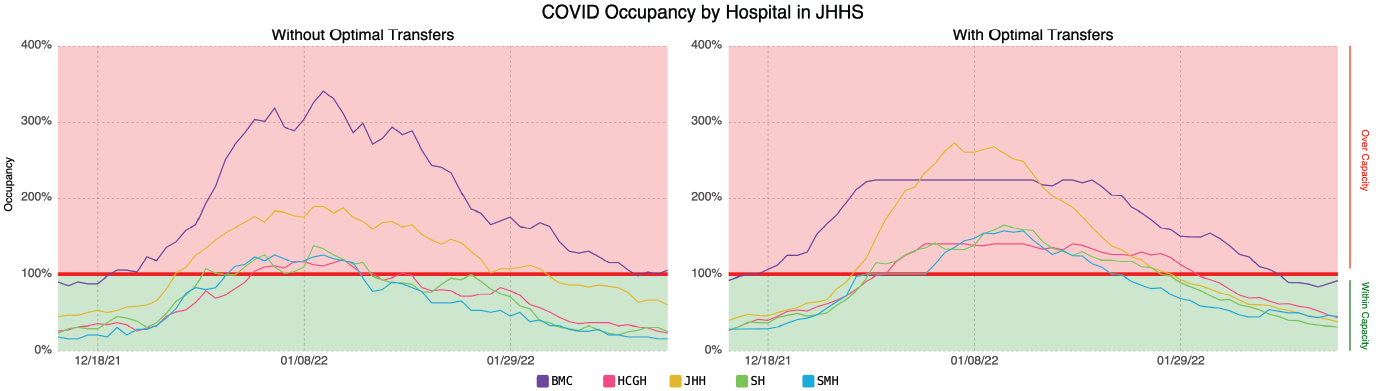}
        \caption{}
        \label{fig:visualizations:load:b}
    \end{subfigure}

    \caption{Screenshots of the visualizations for system and hospital occupancy rates over time, with and without transfers. (a) Shows the past and forecasted occupancy level of the hospital system for COVID-19 patients, relative to the baseline capacity. It comes with a drop-down to select the capacity level occupancy is measured relative to. (b) Plots the occupancy level of each hospital relative to baseline capacity (or other level) for each hospital.}
    \label{fig:visualizations:load}
\end{figure}

The occupancy rate visualizations (\cref{fig:visualizations:load}) provide a high-level view of the occupancy of the hospital system and individual hospitals relative to different capacity levels. The key design feature here is the ability to select the capacity level against which the occupancy is measured. This was included based on user feedback indicating that they needed to assess the system's performance against different capacity thresholds. \Cref{fig:visualizations:load:b} breaks down the occupancy by individual hospital, which helps users quickly identify which hospitals are under the most strain.

\begin{figure}[htb]
    \centering
   \includegraphics[width=0.6\linewidth]{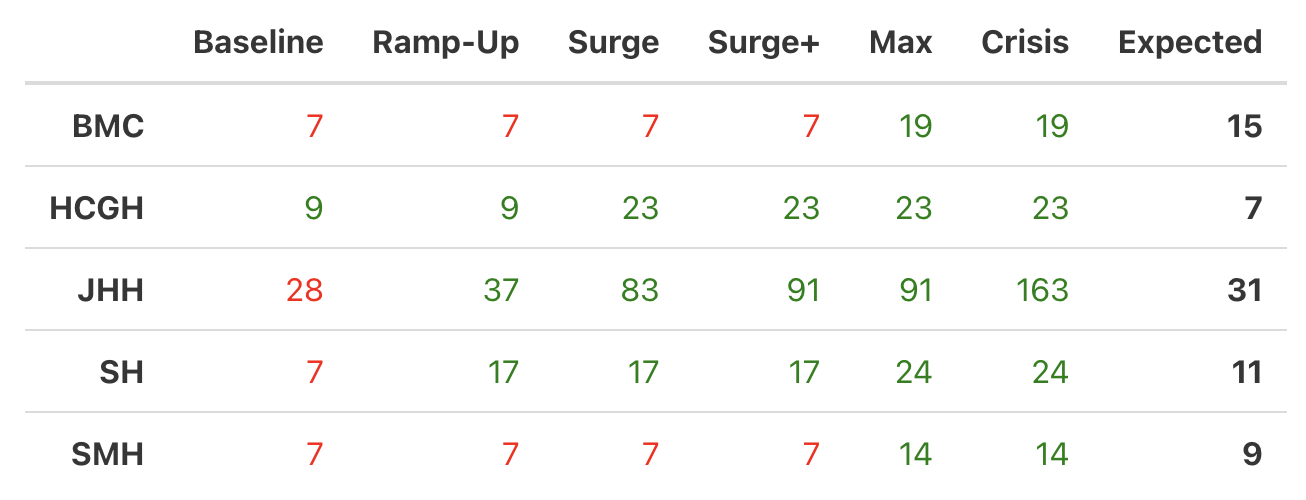}
    \caption{Screenshot of the weekly admission target estimates by hospital and surge level. It provides approximate guidelines on the maximum number of patients each hospital can admit per week at each surge level. It also lists the recent average number of weekly admissions in the last column. Any values below the recent average are shaded red as the corresponding hospital must reduce admissions to reach that level, while all others are green.}
    \label{fig:visualizations:targets}
\end{figure}

Finally, \cref{fig:visualizations:targets} was designed to provide actionable information to users in a format that is immediately understandable and can directly inform operational decisions. It presents admission targets for each hospital to stay under each surge level, along with the recent average admissions. The color-coding allows users to quickly identify when admission levels are unsustainable, indicating that action needs to be taken to avoid exceeding capacity.

One of the main challenges in designing these visualizations was presenting the large amount of complex data in a way that was easy to interpret but still provided sufficient detail for operational decision-making. This was addressed through close collaboration with end-users throughout the design process to ensure the visualizations met their specific needs. Non-standard visualizations, such as the Sankey diagram and the animated map, were used where they could more effectively communicate the information. Interactivity, such as the hover functionality and the capacity level dropdown, was added to allow users to explore the data at different levels of detail without overwhelming them with information.

In summary, the visualizations in the decision-support dashboard were carefully designed to work together to provide users with a comprehensive, easy-to-understand view of the current and predicted state of the hospitals and the impact of the recommended actions. By tailoring the visualizations to the specific needs of the users and using non-standard designs where appropriate, the dashboard enables users to quickly gain the insights they need to make informed decisions about capacity management and patient transfers. The close collaboration with end-users throughout the design process was critical to ensuring that the visualizations were effective in supporting operational decision-making.

\subsubsection{Data Exploration Page}

\begin{figure}[htpb]
    \centering
    \includegraphics[width=.85\textwidth]{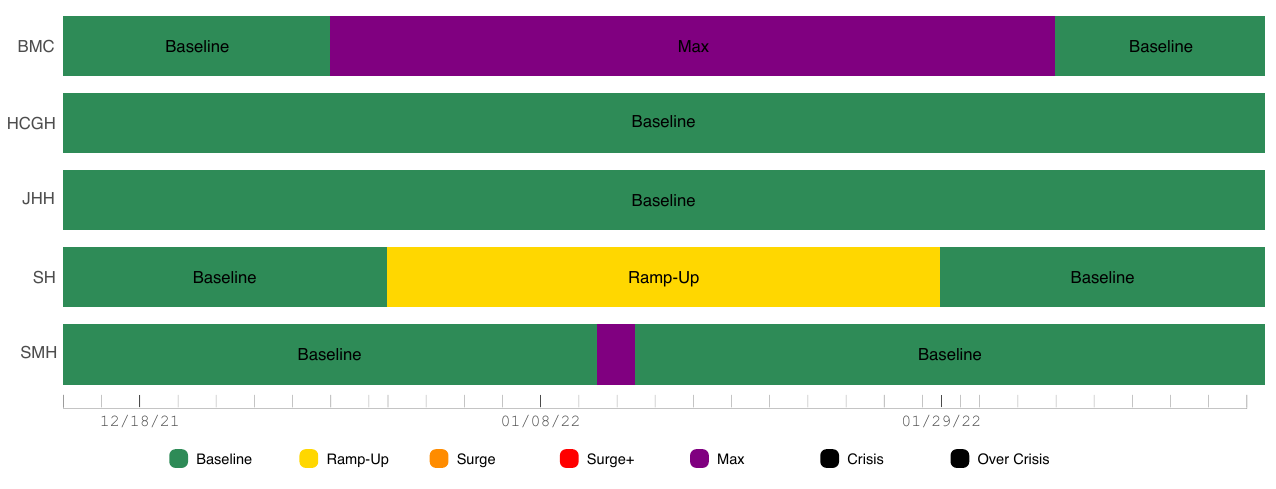}
    \caption{Timeline of surge capacity levels required by each hospital to cope with the projected COVID-19 patient load between December 15, 2021 and February 15, 2022.}
    \label{fig:surge-timeline}
\end{figure}

Recommending decisions to hospital administrators is not the only way to help them make informed capacity management decisions. Visualizations of the relevant data and some key metrics can also help users understand the past, current, and expected load in each hospital which may be valuable in the decision-making process. In particular, the data exploration page shows a line plot of the timeline of the census of each hospital, including both the retrospective data and prospective forecast, along with the capacity levels of the hospital. It allows users to select the patient population and forecast scenarios they are interested in. It also displays an extended version of the surge level timeline seen in \cref{fig:surge-timeline}.

\subsubsection{Status Report}

In addition to the primary decision recommendation tool and the data exploration page, our dashboard includes a status report for the current state of the hospital system to provide an overview of the data and recommended decisions. It summarizes the information on both the decision-support page and the data exploration page and collects key details most relevant to users. This information includes the expected surge level timeline for each hospital, the number of admissions that each hospital can afford on average to stay at their current surge level, and a summary of the recommended transfers.

\subsection{Implementation}
\label{sec:methods:implementation}

The dashboard is implemented as a website with two main components, the backend and the frontend. The backend handles requests for data and the results of our decision-support model with the user’s selected parameters. It is responsible for processing the data, generating the optimization model, solving the model, and returning the results. The backend is written in Julia (v1.6) and uses JuMP (v0.23.2) and Gurobi (v9.5.1) for modeling and solving the optimization model, respectively \citep{Julia-2017, lubin2022, gurobi}. The frontend allows users to select parameters, makes requests to the backend, and displays the results. The figures are generated dynamically on the frontend in JavaScipt using the D3.js library (v7.4) \citep{bostock2011}. The backend runs on a secure server within the JHHS network, and the frontend can be accessed from computers inside this network with a password for security. The code (excluding details specific to JHHS) is made publicly available.\footnote{URL omitted during the review process.}

%% Results %%
\section{Results}
\label{sec:results}

\subsection{Setting}
For the implementation of our decision-support tool within JHHS, data was collected from the five hospitals in the system between March 1st, 2020 and July 1st, 2022. The collected data consists of the number of patients admitted at each hospital during each day with laboratory-confirmed COVID-19 during their stay, and the total number of such patients who were present at each hospital during each day. Approximately N=10,000 patient admissions meeting these criteria were identified, for a total of $\sim$106,000 patient-days. No patient-level data was collected.

\begin{table}
\centering
{\footnotesize
\begin{tabular}{p{5cm}lll} 
\toprule
\textbf{Hospital Name}         & \textbf{Abbreviation} & \textbf{Hospital Type} & \textbf{Bed Count}  \\
\midrule
Bayview Medical Center         & BMC                   & Teaching               & 420                 \\
Howard County General Hospital & HCGH                  & General                & 225                 \\
Johns Hopkins Hospital         & JHH                   & Academic               & 1,091               \\
Suburban Hospital              & SH                    & Community              & 230                 \\
Sibley Memorial Hospital       & SMH                   & Teaching               & 245                 \\
\bottomrule
\end{tabular}}
\caption{Overview of the five hospitals in JHHS.}
\label{tab:hospitals}
\end{table}

\subsection{Case Study}

In this section, we demonstrate some of the results of running the dashboard with different parameters in detail. The default parameters of the dashboard are selected to balance between transfers and surge capacity deployment. To demonstrate the use of the dashboard we also select three other configurations: performing fewer transfers, performing no transfers, and shifting transfers to the largest hospital, JHH.

\begin{figure}[tbh]
\centering
\includegraphics[width=.9\textwidth, trim={3mm 3.5mm 8mm 7.5mm}, clip]{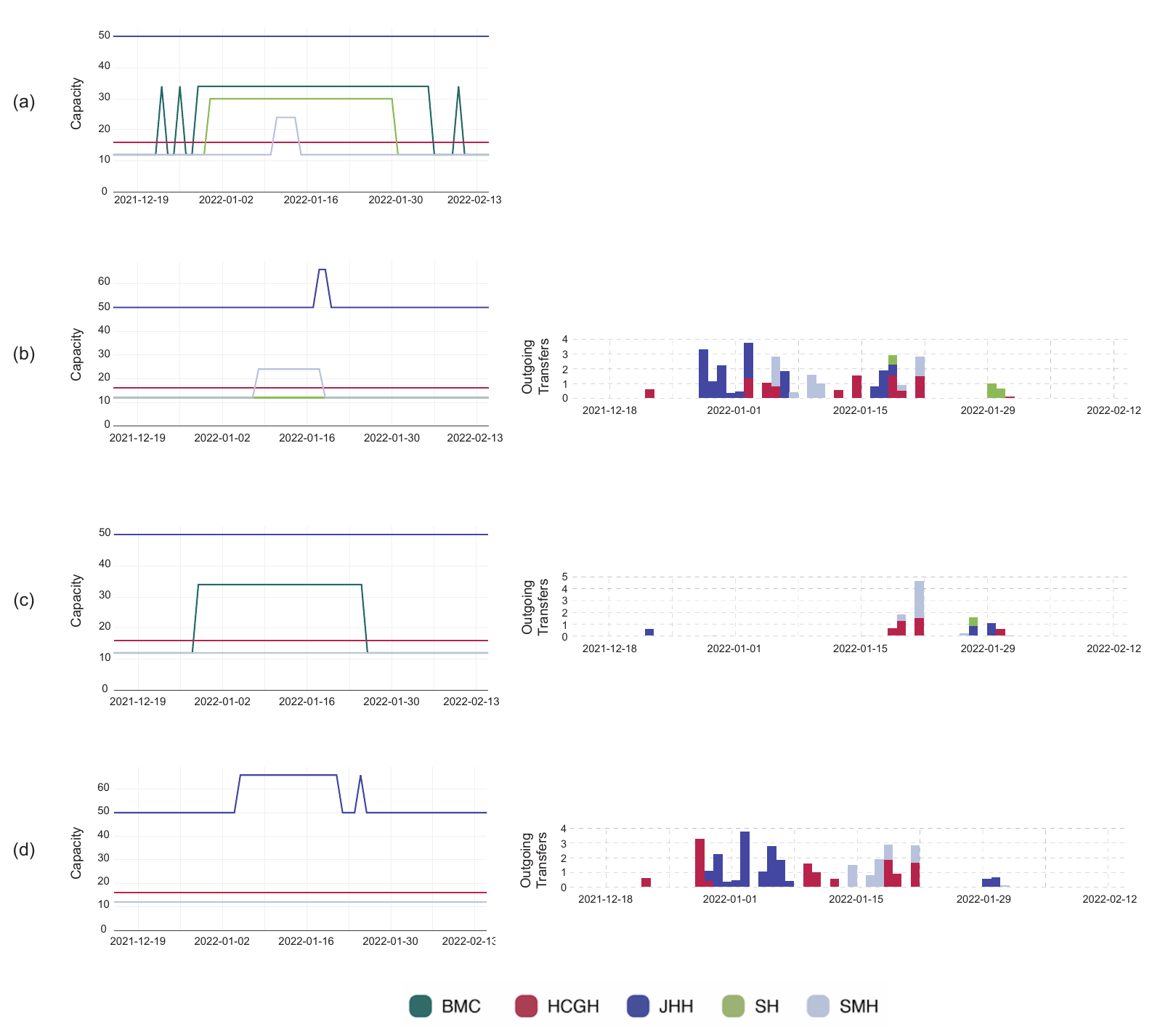}
\begin{subfigure}[b]{0.45\textwidth}
\phantomcaption
\label{fig:casestudy:casestudy-1:a}
\end{subfigure}
\begin{subfigure}[b]{0.45\textwidth}
\phantomcaption
\label{fig:casestudy:casestudy-1:b}
\end{subfigure}
\begin{subfigure}[b]{0.45\textwidth}
\phantomcaption
\label{fig:casestudy:casestudy-1:c}
\end{subfigure}
\begin{subfigure}[b]{0.45\textwidth}
\phantomcaption
\label{fig:casestudy:casestudy-1:d}
\end{subfigure}
\caption{Left: required COVID-19 ICU capacity by hospital in JHHS over a 9-week period of 2021-12-15 to 2022-02-15, assuming discrete capacity levels. Lines for BMC, SH, and SMH frequently overlap as they have the same baseline capacity. Right: recommended outgoing COVID-19 ICU patient transfers from BMC per day. Each bar represents a day, and the color of each section of the bars represents which hospital BMC is sending patients to. Each subplot (a)--(d) represents one of our selected configurations: no transfers, default parameters, fewer transfers, and shift to JHH.}
\label{fig:casestudy:casestudy-1}
\end{figure}

\cref{fig:casestudy:casestudy-1} shows the required dedicated COVID-19 ICU capacity of each hospital between December 15, 2021 and February 15, 2022, and the recommended number of transfers each day from BMC to other hospitals, across four parameter settings.
We focus on the visualization types from \cref{fig:visualizations:capacity:b} and \cref{fig:visualizations:transfers:b} as they display the capacity and transfer recommendations in most detail, respectively.
Without transfers (\cref{fig:casestudy:casestudy-1:a}), BMC and SH frequently have to increase capacity dedicated to COVID-19 ICU patients significantly, which can put them under increased stress and lead to adverse outcomes.
The default parameter setting on the dashboard (\cref{fig:casestudy:casestudy-1:b}) involves approximately 32 transfers from BMC to the other hospitals throughout the two-month period. However, this may be too large of a volume of patients to transfer during such a busy time, particularly since the transfers are so scattered.
If we therefore reduce the transfer budget for BMC, we arrive at the recommendations in \cref{fig:casestudy:casestudy-1:c}. This scenario involves transferring about 10 patients, which may be more practical, and significantly reduces the amount of time BMC needs surge capacity, but does not eliminate it.
We can instead try to make the transfers more consistent and manageable by increasing the relative cost of surge capacity at BMC and decreasing it at JHH, which is the largest hospital in the system and best equipped to handle surges. \Cref{fig:casestudy:casestudy-1:d} shows the impact of this change, which is that most transfers are directed to JHH, and no transfers to SH are needed. The real benefit of this scenario is that all of the surge capacity required is created at JHH, and the other hospitals do not need to expand.
Users can further optimize the recommendations by tweaking other parameters.

\begin{table}[ht]
\centering\footnotesize
\begin{tabular}{llllllll}
\toprule
Hospital & Baseline & Ramp-Up & Surge & Surge+ & Max  & Crisis & Current \\
\hline
BMC  & \cellcolor{red!50}6      & \cellcolor{red!50}6     & \cellcolor{red!50}6   & \cellcolor{red!50}6    & \cellcolor{mygreen}16  & \cellcolor{mygreen}16    & 14       \\
HCGH & \cellcolor{mygreen}7      & \cellcolor{mygreen}7     & \cellcolor{mygreen}19   & \cellcolor{mygreen}19    & \cellcolor{mygreen}19  & \cellcolor{mygreen}19    & 7       \\
JHH  & \cellcolor{red!50}24      & \cellcolor{red!50}32     & \cellcolor{mygreen}73  & \cellcolor{mygreen}80   & \cellcolor{mygreen}80 & \cellcolor{mygreen}143   & 28       \\
SH   & \cellcolor{red!50}6      & \cellcolor{mygreen}14     & \cellcolor{mygreen}14   & \cellcolor{mygreen}14    & \cellcolor{mygreen}21    & \cellcolor{mygreen}21      & 7       \\
SMH  & \cellcolor{mygreen}6      & \cellcolor{mygreen}6     & \cellcolor{mygreen}6   & \cellcolor{mygreen}6    & \cellcolor{mygreen}11  & \cellcolor{mygreen}11    & 0      \\
\bottomrule
\end{tabular}
\caption{Weekly admission targets displayed on the dashboard for each hospital in JHHS at each designated capacity level.}
\label{tab:casestudy:adm-targets}
\end{table}

\Cref{tab:casestudy:adm-targets} provides admission targets for each hospital in the JHHS system at each designated capacity level, along with the recent average admissions, as of December 15, 2021. It mirrors \cref{fig:visualizations:targets}.
These targets represent the approximate number of daily COVID-19 admissions each hospital can sustain while operating at the corresponding capacity level.
For example, if JHH can admit roughly 25 COVID-19 ICU patients per week (that is fewer than 3.5  patients per day on average), it can stay at baseline capacity, but it had recently been admitting 4 COVID-19 ICU patients per day on average, so the corresponding cell in the table is red to indicate that the admission rate was unsustainable. The takeaway for decision-makers would be that they should either consider transfers or increasing the surge level in the future if admission rates continue or increase.
This information enables hospital administrators to proactively manage patient flow and make data-driven decisions about activating surge capacity or implementing measures to reduce admissions when necessary.

\subsection{Workflow}
The decision-support dashboard is designed to be interactive and guide users to iterate toward a solution that matches their needs. The workflow that was developed by end-users of the dashboard began by running the model with the default parameters. The high-level overview figures and statistics were then examined to determine if the recommendations were feasible, if the predicted outcomes met their goals, and if any issues arose, e.g., too many transfers or too much required surge capacity at a hospital. The users could go back and adjust the parameters to address these issues. The model typically runs in 1-10 seconds so users can iterate quickly and efficiently. Once the overview was deemed to be acceptable, users could examine visualizations of interest and make further adjustments. Users would also repeat this process for multiple forecast scenarios to identify adaptable strategies. The figures and numerical results then could be easily downloaded – or the whole page could be saved – for later use. A configuration file containing the exact parameters used could also be saved and uploaded later so that the results and figures could be easily reproduced.

\subsection{Dashboard Usage}
\begin{figure}[htpb]
    \centering
    \includegraphics[width=.8\textwidth]{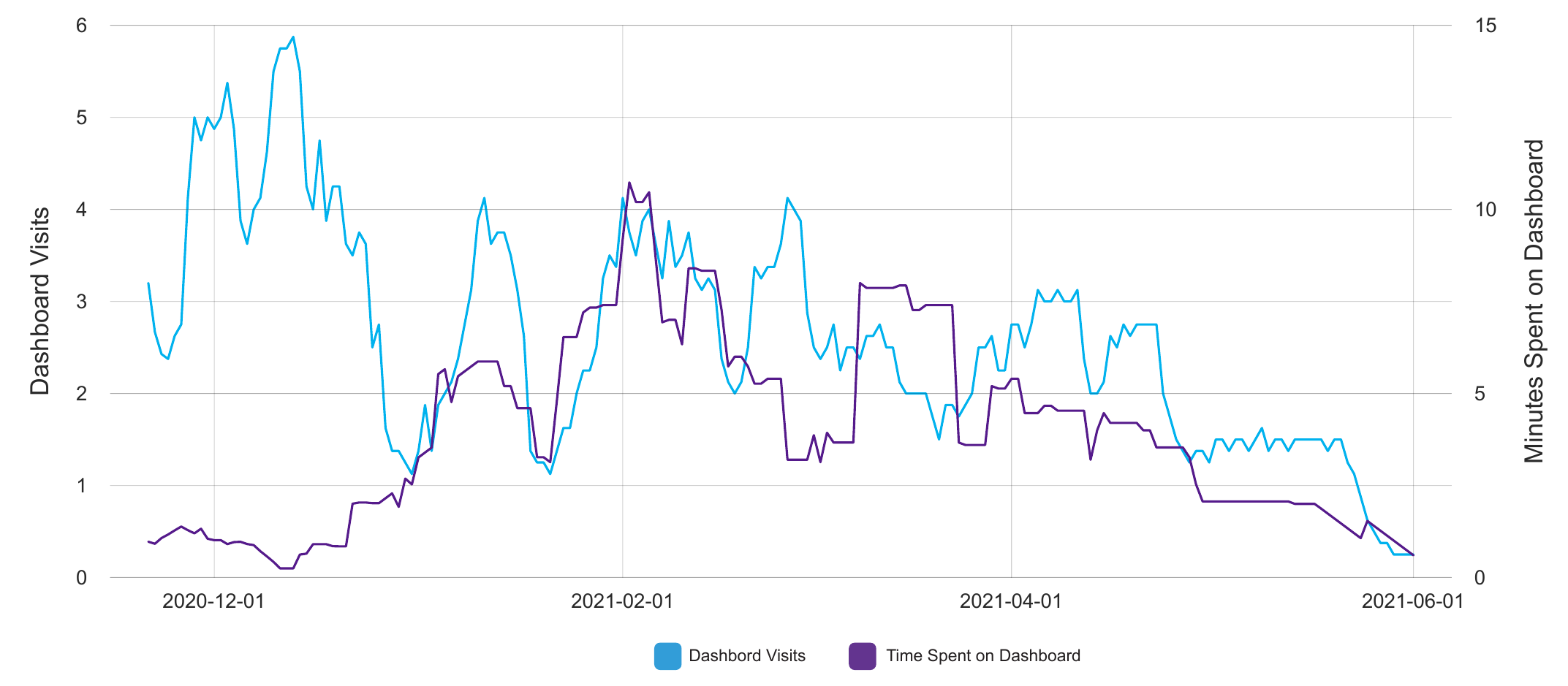}
    \caption{Dashboard visits and the mean amount of time spent on the dashboard by users over time (7-day moving average).}
    \label{fig:dashboard-use}
\end{figure}

The dashboard, actively utilized between November 2020 and June 2021 with over 500 visits, served as a critical tool for forecasting hospital capacity needs during the COVID-19 pandemic. This period, marked by significant COVID-19 surges in May 2020, December 2020, and December 2021 in the Baltimore-Washington, DC region, saw intensive use of the dashboard, particularly in the weeks leading up to these peaks. \Cref{fig:dashboard-use} delineates these usage trends, correlating with the increased demand for model outputs during these critical periods. Two primary end-users accessed the dashboard several times weekly, spending an average of 4 minutes per visit and running the model approximately 2.8 times per visit.

The utility of the dashboard extended beyond individual use, influencing broader decision-making processes. Findings and general results derived from the dashboard were integral in weekly meetings with over 50 stakeholders across the hospital system including the President and Chief Operating Officer of JHHS, hospital administrators, and department chairs. These meetings facilitated the integration of stakeholder feedback into subsequent dashboard iterations, enhancing its relevance and efficacy in managing pandemic-related surges.

%% Discussion %%
\section{Discussion}
\label{sec:discussion}

For a decision-support tool to be useful in practice, it must aim for trustworthiness, transparency, explainability, usability, and respectful implementation. We approached these challenges with our participatory design approach that included all hospital capacity stakeholders, including administrators and clinicians, in the development cycle. Transparency and trustworthiness were achieved by training and working with hospital data scientists on the inner workings of the tool, including data sources and optimization model formulation. We increased trustworthiness in our recommendations by making sure model inputs – data and parameters – were complete and accurate against official hospital records and stakeholders’ beliefs. Explainability was achieved by incorporating all feedback from stakeholders and made major changes to ensure our dashboard was usable and implemented responsibly.
 
 As an example, one shortcoming of the initial model was that it assumed the burden of increasing capacity or transferring patients was similar across all hospitals instead of depending on the size and capabilities of the individual hospitals. The objective function was therefore modified to allow users to adjust the burden of adding capacity or transferring patients, enabling users to impose a preference of creating capacity at or transferring to larger, more capable hospitals, or to load-balance patients evenly to increase fairness.

The parameters that users have control over were selected through a participatory design process. The workflow evolved such that analysts used the tool to generate and present results to stakeholders, instead of stakeholders directly using the dashboard. This allowed us to enhance the tool with more parameters and control for the user, without making it overly complex. We worked with the end-users to explain each parameter and its effects, and added concise explanations to the dashboard through hoverable pop-ups. Stakeholders prioritized surge capacity deployment over patient transfers, so customizable transfer limits were added.

The metrics and visualizations presented were also selected and modified through the participatory design process. We found that in general, the users and stakeholders were more focused on the projected impacts of decisions rather than the details of the decisions themselves, so additional visualizations of the predictions were created. Stakeholders preferred high-level, easily interpretable visualizations, such as \cref{fig:surge-timeline}, over more detailed but unfamiliar visualizations, so those were emphasized and improved. Certain interpretable metrics such as admission targets were also prioritized. However, detailed figures and statistics were still useful for analysts, so we created individual sections for each analysis that could be easily shown or hidden depending on the audience, rather than removing these details.

This work has some limitations and opportunities for future research. One limitation is that it requires knowledge and experience to be able to interpret and use effectively, which can be a significant barrier to adoption. More work can be done to improve user experience, interface, and figures, as well as providing more information to help users understand the insights. Additional analytics on the usage of the dashboard and quantitative measures of its impact would provide more value in refining it further. Finally, like any data-driven modeling approach, our dashboard relies on data and forecasts that may be unavailable or inaccurate.

%% Conclusion %%
\section{Conclusion}
\label{sec:conclusion}

In summary, we developed a practical, valuable tool for improving capacity management decisions in hospital systems during periods of high demand. We built an interactive dashboard to support decision-making that uses real-time data and forecasts to drive a prospective customizable optimization model that recommends optimal decisions in a data-driven manner, guided by the user’s preferences. This tool provides hospital administrators with actionable insights into the potential impacts of the decisions they make for pandemic surges or other events. It is flexible and can be fine-tuned to the specific goals, constraints, and operational realities of the hospitals.

%% Bibliography %%
%\newpage
\singlespacing
\bibliographystyle{elsarticle-num-names}
\bibliography{jhhs-dashboard}

%% Appendix %%
\newpage
\appendix

\section{Complete Optimization Model}
\label{appendix:model}
\singlespacing

For the sake of completeness, in this appendix we detail the full capacity management optimization model which was discussed in \cref{sec:methods:model}. The relevant notation is described in \cref{tab:appendix:notation}.
See \citet{parker2024cm} for additional discussion of this model.

\begin{table}[ht]
	\centering\footnotesize
	\begin{tabular}{p{1.8cm}p{3.5cm}p{10cm}}
		\toprule
		\textbf{Notation	}		& \textbf{Name}		& \textbf{Description}		\\
		\midrule
		\multicolumn{3}{c}{\textit{Sets}} \\
		\midrule
		\( \T \)		& Time steps		& Set of all time steps (days)		\\
		\( \H \)		& Hospitals		& Set of hospitals		\\
		\( \C \)		& Bed groups		& Set of bed groups or units of capacity		\\
		\midrule
		\multicolumn{3}{c}{\textit{Data}} \\
		\midrule
		\( i_{h,t} \)		& Incoming demand		& Number of patients arriving at hospital $h$ during time period $t$		\\
		\( b_{h,k} \)		& Capacity per unit		& Number of staffed beds at hospital $h$ in unit $k$		\\
		\( L_h \)		& Length of stay		& Distribution over length of stay for patients at hospital $h$		\\
		\midrule
		\multicolumn{3}{c}{\textit{Decision Variables}} \\
		\midrule
		\( u_{h,t,k} \)		& Unit usage		& Binary variable indicating whether unit $k$ at hospital $h$ is in use during time period $t$		\\
		\( s_{h,g,t} \)		& Patient transfers		& The number of patients to transfer/divert from hospital $h$ to hospital $g$ during time period $t$		\\
		\midrule
		\multicolumn{3}{c}{\textit{Auxillary Variables}} \\
		\midrule
		\( o_{h,t} \)		& Census		& Estimated census of hospital $h$ at time $t$		\\
		\( a_{h,t} \)		& Admissions		& Recommended admissions (including transfers) at hospital $h$ at time $t$		\\
		\( d_{h,t} \)		& Discharges		& Estimated discharges from hospital $h$ at time $t$		\\
		\( c_{h,t} \)		& Allocated capacity		& Operational capacity allocated at hospital $h$ during time period $t$		\\
		\( \hat{u}_{h,t,k} \)		& Unit availability		& Binary variable indicating whether unit $k$ at hospital $h$ is available for the target population during time period $t$		\\
		\( \check{u}_{h,t,k} \)		& Unit conversion		& Binary variable indicating whether unit $k$ at hospital $h$ was converted during time period $t$		\\
		\midrule
		\multicolumn{3}{c}{\textit{Parameters}} \\
		\midrule
		\( w^{(1)}, w^{(2)}, w^{(3)} \)		& Objective function weights		& Costs or weights for the optimization objective function		\\
		\( S \)		& Transfer limits		& Upper bounds on the number of patients that can be transferred		\\
		\( \Gamma_1, \Gamma_2 \)		& Uncertainty set parameters		& Control the deviation of uncertain scenarios from the nominal predictions	\\
		\( \delta^+_{h,k}, \delta^-_{h,k} \)		& Setup time		& Number of time periods to convert capacity to/from the target population	\\
		\bottomrule
	\end{tabular}
	\caption{Notation used in the complete optimization model.}
	\label{tab:appendix:notation}
\end{table}

\small
% objective function
\begin{subequations}
\begin{align}
    \text{min}  \qquad & \sum_{t \in \T} \sum_{h \in \H} \sum_{k \in \C} \left( w^{(1)}_{h,t} c_{h,t} + w^{(2)}_{h,k} \hat{u}_{h,t,k} + w^{(3)}_{h,k} \check{u}_{h,t,k} \right)
	\label{ap:eq:obj}\\
%\end{equation}
%
% census calculation from admissions
%\begin{subequations}
	&o_{h,t} = \sum_{t^\prime = 1}^{t} \left( a_{h,t^\prime} - d_{h,t^\prime} \right)		&\forall h \in \H, t \in \T		\label{ap:eq:exp:census} \\
	&a_{h,t} = i_{h,t} + \sum_{g \in \H} \left( s_{g,h,t} - s_{h,g,t} \right)			&\forall h \in \H, t \in \T		\label{ap:eq:exp:admissions} \\
	&d_{h,t} = \sum_{t^\prime = 1}^{t} \left( P(L_h = t - t^\prime) a_{h,t^\prime} \right)		&\forall h \in \H, t \in \T	\label{ap:eq:exp:discharges}\\
%\end{align}
%\end{subequations}
%
% capacity
%\begin{subequations}
%\begin{align}
	&c_{h,t} = \sum_{k \in \mathcal{C}} u_{h,t,k} b_{h,k}		&\forall h \in \H, t \in \T	\label{ap:eq:exp:capacity} \\
	&o_{h,t} \leq c_{h,t}			&\forall h \in \H, t \in \T	\label{ap:eq:cons:no-shortage} \\
	&u_{h,t,k} \in \set{0,1}			&\forall h \in \H, t \in \T, k \in \C
\end{align}
\end{subequations}

\textbf{Optional constraints:}

Variable definitions:
% capacity allocated (available or not)
\begin{subequations}
\begin{align}
	u_{h,t,k} &\leq \hat{u}_{h,t-\delta^-_{h,k},k}		&\forall h \in \H, t \in \T, k \in \C	\label{ap:eq:cons:allocated:a} \\
	u_{h,t,k} &\leq \hat{u}_{h,t+\delta^+_{h,k},k}		&\forall h \in \H, t \in \T, k \in \C	\label{ap:eq:cons:allocated:b} \\
	u_{h,t,k} &\leq \hat{u}_{h,t,k}		&\forall h \in \H, t \in \T, k \in \C	\label{ap:eq:cons:allocated:c} \\
	\hat{u}_{h,t,k} &\in \set{0,1}			&\forall h \in \H, t \in \T, k \in \C	\label{ap:eq:cons:allocated:d}\\
%\end{align}
%\end{subequations}
%
% indicator for capacity conversion
%\begin{subequations}
%\begin{align}
	\check{u}_{h,t,k} &\geq u_{h,t,k} - u_{h,t-1,k}		&\forall h \in \H, t \in \T \setminus \set{1}, k \in \C	\label{ap:eq:cons:conversion:a} \\
	\check{u}_{h,1,k} &= u_{h,1,k}		&\forall h \in \H, k \in \C	\label{ap:eq:cons:conversion:b} \\
	\check{u}_{h,t,k} &\in \set{0,1}			&\forall h \in \H, t \in \T, k \in \C	\label{ap:eq:cons:conversion:c}
\end{align}
\end{subequations}

Constraints:
% allocate in order, maximum utilization, transfer limits
\begin{subequations}
\begin{align}
    &u_{h,t,k} \leq u_{h,t,k-1}  &\forall h \in \H, t \in \T, k \in \C
	\label{ap:eq:cons:unit-order}\\
    \
    &o_{h,t} \leq z c_{h,t}		&\forall h \in \H, t \in \T		\label{ap:eq:cons:util-rate-1} \\
	&o_{h,t} \leq c_{h,t} - z^\prime		&\forall h \in \H, t \in \T		\label{ap:eq:cons:util-rate-2} \\
	&s_{h,g,t} \leq S_{h,g}		&\forall h \in \H, g \in \H, t \in \T		\label{ap:eq:cons:transfer-limits:1} \\
	&\sum_{g \in \H} s_{h,g,t} \leq S_{h}		&\forall h \in \H, t \in \T		\label{ap:eq:cons:transfer-limits:2} \\
	&\sum_{h,g \in \H} \sum_{t \in \T} s_{h,g,t} \leq S		&	\label{ap:eq:cons:transfer-limits:3}
\end{align}
\end{subequations}

\textbf{Uncertainty sets:}

% uncertainty set for census only
%\begin{equation}
%	O_h = \left\{  o_h \in \mathbb{R}^{|\T|} \big\vert o_{h,t} \in [\bar{o}_{h,t} - \tilde{o}^-_{h,t}, \bar{o}_{h,t} + \tilde{o}^+_{h,t}], 
%	\frac{\sum_{t \in \T} \left| o_{h,t} - \bar{o}_{h,t} \right|}{\sum_{t \in \T} \bar{o}_{h,t}} \leq \Gamma_1, 
%		\left| \frac{o_{h,t} - o_{h,t+1}}{\bar{o}_{h,t} - \bar{o}_{h,t+1}} \right| \leq \Gamma_2, \quad\forall t \in \T
%        \right\}
%	\label{eq:exp:uncertainty-set-census}
%\end{equation}

% uncertainty set for census based on admissions
\begin{subequations}
\begin{align}
	I_h &= \left\{ 
             i_h \in \mathbb{R}^{|\T|} :
             i_{h,t} \in [\bar{i}_{h,t} - \tilde{i}^-_{h,t}, \bar{i}_{h,t} + \tilde{i}^+_{h,t}], 
		 \frac{\sum_{t \in \T} \left| i_{h,t} - \bar{i}_{h,t} \right|}{\sum_{t \in \T} \bar{i}_{h,t}} \leq \Gamma_1,
		\left| \frac{i_{h,t} - i_{h,t+1}}{\bar{i}_{h,t} - \bar{i}_{h,t+1}} \right| \leq \Gamma_2, \quad\forall t \in \T
         \right\}
	\label{ap:eq:exp:uncertainty-set-incoming}\\
    \
	O^\prime_h &= \left\{ \tilde{o}(i_h, s, \mathcal{L}_h) : i_h \in I_h \right\}
	\label{ap:eq:exp:uncertainty-set-census-prime}
\end{align}
\end{subequations}

\end{document}